\documentclass[a4paper,11pt]{article}

\usepackage{jheppub} 

\usepackage{lmodern}
\usepackage{verbatim}
\usepackage{amsmath,amssymb}
\usepackage{hyperref}
\usepackage{xcolor}

\usepackage{graphicx}

\usepackage{tikz}
\usetikzlibrary{positioning,decorations.pathmorphing}

\let\oldFootnote\footnote
\newcommand\nextToken\relax

\renewcommand\footnote[1]{%
    \oldFootnote{#1}\futurelet\nextToken\isFootnote}

\newcommand\isFootnote{%
    \ifx\footnote\nextToken\textsuperscript{,}\fi}

\usepackage{enumerate}

\usetikzlibrary{decorations.markings}

\def\id{{1 \kern-.28em {\rm l}}}

\def\K3{{\bf K3}}
\def\journal#1&#2(#3){\unskip, \sl #1\ \bf #2 \rm(19#3) }
\def\andjournal#1&#2(#3){\sl #1~\bf #2 \rm (19#3) }

\def\bar{\overline}

\def\ie{{\it i.e.}}
\def\eg{{\it e.g.}}

\def\etc{{\it etc}}

\def\tilde{\widetilde}

\def\frac#1#2{{#1\over#2}}

\def\inbar{\,\vrule height1.5ex width.4pt depth0pt}
\def\IC{\relax\hbox{$\inbar\kern-.3em{\rm C}$}}
\def\IR{\relax{\rm I\kern-.18em R}}
\def\IP{\relax{\rm I\kern-.18em P}}

\catcode`\@=11
\def\slash#1{\mathord{\mathpalette\c@ncel{#1}}}
\def\underrel#1\over#2{\mathrel{\mathop{\kern\z@#1}\limits_{#2}}}

\catcode`\@=12

\def \sinh{{\rm sinh}}

\def\exp{{\rm exp}}

\def\ie{{\it i.e.}}
\def\eg{{\it e.g.}}

\title{ Holographic Complexity of LST and Single Trace $T\bar{T}$}

\author{Soumangsu Chakraborty$^a$,  Gaurav Katoch$^b$, Shubho R. Roy$^b$}
\emailAdd{soumangsuchakraborty@gmail.com}
\emailAdd{katoch.gaurav1@gmail.com}
\emailAdd{roy.shubho@gmail.com}

\affiliation{$^a$Department of Theoretical Physics,\\Tata Institute of Fundamental Research,\\Homi Bhabha Road, Mumbai 400005, India}
\affiliation{$^b$Department of Physics,\\Indian Institute of Technology, Hyderabad\\
Kandi, Sangareddy 502285, Medak, Telengana, India}

\abstract{In this work, we continue our study of string theory in the background that interpolates between $AdS_3$ in the IR to flat spacetime with a linear dilaton in the UV. The boundary dual theory interpolates between a CFT$_2$ in the IR to a certain two-dimensional Little String Theory (LST) in the UV. In particular, we study \emph{computational complexity} of such a theory through the lens of holography and investigate the signature of non-locality in the short distance behavior of complexity. When the cutoff UV scale is much smaller than the non-locality (Hagedorn) scale, we find exotic quadratic and logarithmic divergences (for both volume and action complexity) which are not expected in a local quantum field theory. We also generalize our computation to include the effects of finite temperature. Up to second order in finite temperature correction, we do not any find newer exotic UV-divergences compared to the zero temperature case. }

\begin{document}
\maketitle
\flushbottom

\section{Introduction}

AdS/CFT \cite{Maldacena:1997re, Gubser:1998bc, Witten:1998qj, Aharony:1999ti}, or more broadly speaking gauge/gravity duality \cite{Itzhaki:1998dd}, has revolutionized our understanding of strongly coupled quantum field theories. For a large class of field theories, calculations which were once considered beyond reach due to breakdown of coupling constant perturbation theory are now routinely being done by first mapping the field theory to its gravity dual (often constructed from the ``\emph{bottom up}" without even the need for knowledge of any details of string theory), and then solving (numerically in most cases) the classical gravity-matter system, \ie\ Einstein field equations coupled to classical matter fields. This so called ``\emph{holographic approach}" of solving strongly coupled (gauge) fields theories have extended the use of gravitational methods (GR/SUGRA) to the fields of condensed matter physics \cite{Sachdev:2010ch, McGreevy:2009xe, Hartnoll:2009sz} and QCD \cite{Erlich:2005qh, DaRold:2005mxj, Karch:2006pv}. However, the impact of AdS/CFT (gauge/gravity)  has been far more deep and revealing than merely providing a classical geometrical computational tool for strongly coupled field theory phenomena. Thinking about how field theory codes various phenomena on the gravity side, such as emergence of a quasilocal bulk spacetime local observables propagating on it, spatial connectivity of the bulk geometry, event horizons and gravitational singularities \etc, has led to the recognition and importance of various concepts from the quantum information and computation (QIC) literature which capture aspects of quantum field theories not captured by traditional observables such as correlation functions of local operators or Wilson loops. Information geometry/information metrics, Von-Neumann \cite{Ryu:2006bv, Hubeny:2007xt} and Renyi \cite{Dong:2016fnf} Entropy, Mutual Information, Tensor networks \cite{Swingle:2009bg}, Computational Complexity, Fidelity susceptibility, Quantum error correcting codes are only to name a few. This has become a highly productive enterprise leading to insights which might even solve the information paradox \cite{Penington:2019npb, Almheiri:2019hni}. Combining insights from holographic gravity duals, from integrability or supersymmetry based arguments, from lattice based approaches and perturbative approaches, we have explored the landscape of local quantum field theories rather comprehensively. However, the landscape of nonlocal quantum field theories is still mostly \emph{terra incognita}. 

In this work, we focus our attention on the decoupled theory on a stack of $k\gg1$ NS5 branes wrapping $T^4 \times S^1$, the so called Little String theory (LST) in $1+1$ dimensions. Unlike Dp branes,  the worldvolume theory living on the NS5 branes decouples from the bulk  at finite value of the string length $l_s=\sqrt{\alpha'}$. This is a signature that the decoupled theory namely LST living on the NS5 branes is not a local field theory.  In fact the decoupled theory on the NS5 branes is somewhat intermediate between string theory (which is not a local theory and gives rise to massless gravitons upon quantization) and a local quantum field theory. The holographic background obtained by taking the near horizon geometry of the NS5 branes is flat spacetime with a linear dilaton $\mathbb{R}^{1,1}\times \mathbb{R}_\phi$. Such a holographic duality has been studied quite extensively in \cite{Aharony:1998ub,Kutasov:2001uf}.

Next, let us introduce $p\gg 1$ F1 strings wrapping the $S^1$. The near horizon geometry of the F1 strings is given by $AdS_3$. Thus the full geometry interpolates between $AdS_3$ in the IR (which corresponds to the near horizon geometry of the F1 strings) to flat spacetime with a linear dilaton in the UV (which corresponds to the near horizon geometry of just the NS5 branes). Correspondingly, the boundary field theory interpolates between  a local CFT$_2$ dual to $AdS_3$ in the IR to LST in the UV. The interpolating geometry discussed above is often referred to in the literature as $\mathcal{M}_3$.

After the advent of $T\bar{T}$ deformation \cite{Smirnov:2016lqw,Cavaglia:2016oda}, it was realized in \cite{Giveon:2017nie} that there is  a deformation of string theory in $AdS_3$ that shares many properties in common with the  double trace $T\bar{T}$ deformation.\footnote{Details of the single trace $T\bar{T}$ deformation appear in section \ref{sec2}.}  Such a deformation, often referred to in the literature as the single trace $T\bar{T}$ deformation,  of string theory in $AdS_3$, changes the UV asymptotics of the bulk geometry from $AdS_3$ to flat spacetime with a linear dilaton keeping fix the IR regime of the geometry. Analysis in \cite{Giveon:2017nie} shows that the dual background geometry interpolates between $AdS_3$ in the IR to flat spacetime with a linear dilaton in the UV.   Holography in this background (often referred to as $\mathcal{M}_3$) can be realized as a concrete example of holography in non-AdS background that is smoothly connected to $AdS_3$.

This work is a part of a growing body of literature over the last few years to employ holography to investigate various aspects of nonlocal field theories such as LST which admit gravity duals \cite{Asrat:2017tzd,Chakraborty:2018kpr,Chakraborty:2018aji,Chakraborty:2020xyz,Chakraborty:2020udr,Chakraborty:2020yka}. We are optimistic that holography will be as productive in demystifying properties of nonlocal quantum field theories such as the LST as it has been for enhancing our understanding of strongly coupled regimes of local field theories. Our current understanding of holography is that the bulk spacetime represents an encoding of entanglement structure \ie\ quantum mechanical correlations of the dual field theory  degrees of freedom/state \cite{VanRaamsdonk:2009ar, VanRaamsdonk:2010pw}. The famous Ryu-Takayanagi (RT) proposal \cite{Ryu:2006bv, Hubeny:2007xt} was one of the earliest major piece of evidence to point in this direction (along with Maldacena's construction \cite{Maldacena:2001kr} of the eternal Schwarzschild-AdS (SAdS) as a thermally entangled state of two CFTs). Since then a long impressive list of quantum entanglement related CFT structures/observables have been related to classical geometric features of the bulk (see \eg\  \cite{VanRaamsdonk:2016exw} for a review). However, entanglement entropy or related concepts such as tensor networks or error-correcting codes are still unable to capture the essential features of bulk geometry which are masked behind the black hole horizons. Take for example the growth of the Einstein-Rosen Bridge (ERB) behind the horizon. Entanglement entropy saturates in a short time upon reaching thermalization whereas, ER bridge continues to grow linearly with time long after the system hits thermalization. To explain the ERB growth, Susskind  \cite{Susskind:2014rva} has recently borrowed another tool from quantum information theory and added to the holographic dictionary, namely the \emph{computational complexity}. Complexity is the quantity associated with the states in the Hilbert space of the field theory living on the boundary which quantifies the difficulty of preparing a state (called the target state), starting from the given reference state. This is a well defined quantity for discrete systems, like quantum circuits in information theory. But it turned out to be hard to define complexity for the continuous systems described by QFT, where the precise definition of complexity is still lacking. To cope with this ambiguity, Nielsen et. al.\cite{2005quant.ph..2070N, 2006Sci...311.1133N} provide a definition of circuit complexity in
field theory as the minimum number of unitary gates in the space of unitary operators which has a Finsler geometry. The complexity of a target state, given a reference state, is defined as the geodesic length in Finsler manifold with suitable cost functions which acts like Lagrangian in typical variational problem. These cost functions are required to obey certain desirable conditions like continuity, positive definiteness and triangle inequality \etc. Despite this attempt at achieving precision, there is still arbitrariness in the choice of cost functions which fixes the Finsler metric and complexity depends upon the choice of the metric. Several attempts have been made to define complexity in the continuum limit (see \eg\ \cite{Jefferson:2017sdb, Chapman:2017rqy, Khan:2018rzm, Yang:2018nda, Molina-Vilaplana:2018sfn, Hackl:2018ptj, Bhattacharyya:2018wym, Guo:2018kzl, Bhattacharyya:2018bbv, Yang:2018tpo, Camargo:2019isp, Balasubramanian:2019wgd, Bhattacharyya:2019kvj, Erdmenger:2020sup, Bueno:2019ajd, Chen:2020nlj, Flory:2020eot, Flory:2020dja} for an incomplete but representative list). However, it is fair to say that up to now there exists neither any universal and/or unanimous definition of complexity in the continuum limit nor an exhaustive study of its possible universality classes. In particular, in the continuum, complexity, even  in principle, is a UV divergent quantity because it is defined to within a tolerance ($\epsilon$) with respect to the target state. Demanding more precision of replicating the target state requires insertion of more number of gates which leads to a dependence on the inverse tolerance which is a divergent term. A similar trend emerges from the bulk perspective as both the definitions involve the integrations over infinite regions of spacetime (bulk IR divergence). Usually, the divergent or explicitly cutoff dependent quantities in QFT are not considered physical as their value can be changed by changing the UV cutoff. But this characteristic UV dependence of complexity is a feature which seems to be relaxed while defining complexity
in QFT. 

There are two proposals in holography, each with its own merits and motivation, as to what is the bulk geometric dual of the
complexity of a boundary field theory quantum state. One is that the field theory complexity should be proportional to the volume of the maximal volume spatial surface extending into the bulk and terminating on the boundary at the spacial slice on which the boundary quantum state is defined \cite{Susskind:2014rva}. This is referred as the complexity-volume ($CV$) conjecture. The other proposal \cite{Brown:2015bva, Brown:2015lvg} is that the complexity is proportional to the bulk on-shell action integral evaluated or supported in the Wheeler-deWitt (WdW) patch of the boundary spatial slice on which the field theory state is specified\footnote{The WdW patch of a given spatial slice on the boundary is defined to be the bulk subregion covered by the union of all possible spacelike surfaces in the bulk which terminates on the same spatial slice at the
boundary.}. This is called the complexity-action ($CA$) proposal. Both these bulk measures of complexity are manifestly UV divergent, so regularization is necessary as remarked before. In the $CV$ proposal there is an inherent ambiguity - to make the expression dimensionally consistent one must include a \emph{characteristic length scale}, $L$, of the geometry for which there is no unique prescription. For the $CA$ proposal, there are also couple of issues. Some boundaries of the WdW patch are codimension one null/lightlike submanifolds \emph{with} joints/edges. The presence of such null boundaries and their joints (edges) entails that the GHY boundary terms be properly defined as discussed in \cite{Lehner:2016vdi}. In this paper, we take a different approach to this problem \cite{Brown:2015lvg, Parattu:2015gga, Bolognesi:2018ion}. Since we have to UV-regulate the WdW patch anyways, we use a particular regularization which deforms the WdW null boundary to timelike and in the process also smooths out the joints. In this way we can compute the GHY terms in one step without any issues.

The plan of the paper is as follows. In section \ref{sec2}, for the sake of completeness, we give a brief review of string theory in $AdS_3$ and its single trace $T\bar{T}$ deformation and mention certain interesting features of LST. In section \ref{0T complexity}, we set out to compute the holographic complexity of the spacetime theory dual to string theory in $\mathcal{M}_3$, using first the $CV$ prescription and then the $CA$ prescription. However, since the bulk contains a non-trivial dilaton field, we first propose a generalized definition of the volume complexity in the string frame.  This generalized prescription guided by the requirement of furnishing the correct powers of the string coupling \ie\ $G_N$ in the complexity expression\footnote{Similar considerations led the authors in \cite{Klebanov:2007ws} to a generalization of the Ryu-Takayanagi formula for holographic entanglement entropy for bulk backgrounds supporting a non-trivial dilation in the string frame.}. The volume complexity immediately reveals the nonlocal nature of the dual field theory (LST). For a local field theory, extensivity property of the complexity means that leading term in complexity scales with the spatial volume of the dual field theory in units of the lattice cell volume. Thus the leading piece diverges as the inverse lattice cell volume \ie\ $\epsilon^{-d}$ where $d$ is the number of spatial dimensions of the boundary theory and $\epsilon$ is the short distance cutoff. In the present case, the dual field theory has only one spatial dimension ($d=1$), so one would naively expect the volume complexity to diverge as $C_{V}\sim 1/\epsilon$. Here instead the leading piece of complexity diverges \emph{quadratically} with the UV cutoff, $C_{V}\sim \frac{1}{\epsilon^2}$! This behavior is dominant when the $\epsilon/\beta_H\ll1$ where and $\beta_H$ is the inverse Hagedorn temperature of LST can also be thought of as the non-locality scale of LST. When $\epsilon/\beta_H\gg 1$, then one recovers the scaling of complexity with spatial  (for volume complexity), \ie\ of a local quantum field theory. 
  Thus one can conclude that for length scales below $\beta_H$ (the non-locality scale of LST), stringy physics takes over and the theory departs from behaving like a local field theory. As further features of non-locality, we find logarithmic divergent pieces (subleading divergence) in the complexity expression when $\epsilon/\beta_H\ll1$. The dimensionless universal constant which appears as the coefficient of the log divergence can be given the interpretation of the total number of  ``regularized/effective" degrees of freedom in the spacetime theory as opposed to the true degrees of freedom of LST that diverges \cite{Barbon:2008ut,Chakraborty:2018kpr}.
 The action complexity results display the exact same divergence structures, quadratic and logarithm when $\epsilon\ll \beta_H$. Modulo an overall constant (inherent in the ambiguity of the ``characteristic length-scale" in the definition of the volume complexity), the leading quadratic divergence piece matches for both the volume and action complexities. However, it is interesting to note that the subleading logarithmic divergence, while same in magnitude, \emph{differs by a sign} in the volume and action complexity expressions. This is not a novel observation. Past studies have revealed that the coefficients of the subleading divergent pieces might be different \cite{Bolognesi:2018ion}  hinting to the fact that the two bulk/holographic prescriptions of complexity might actually correspond to different schemes of defining complexity in the boundary field theory.  
  As a check, we extract the behavior of the spacetime theory action complexity in the deep IR limit (\ie\ $\epsilon\gg\beta_H$) where it indeed reproduces the pure AdS/CFT vacuum state complexity \cite{Reynolds:2016rvl, Carmi:2016wjl} for both prescriptions.  So far everything we've said here corresponds to the zero temperature case. Since the LST is a nonlocal theory for which we do not have much intuition, there might appear novel exotic divergences compared to the zero temperature case - so it was imperative that we study the finite temperature case. With this aim, in section \ref{finiteT}, we explore the effects of finite temperature in LST. In particular, we consider the thermofield double state of two LST's for which the dual bulk geometry is an eternal $\mathcal{M}_3$ black hole. Since for such a gravity background analytic calculations of the maximal volume slice without any approximations are not possible, we abandon the volume complexity scheme and instead numerically compute the action complexity exactly. The plot of the complexity as a function of $\epsilon/\beta_H$, displayed in figure \ref{img4}. Qualitatively the action complexity at finite temperature exhibits the same behavior as that of the zero temperature case. More importantly, \emph{no new divergences} arise compared to the zero temperature case perturbatively up to second order in finite temperature corrections. Finally, in section \ref{DiscOut} we conclude by discussing our results and provide an outlook for future work.

One shortcoming of both these holographic proposals for evaluating circuit complexity of the boundary theory is that there is no explicit reference to the boundary reference state as well the unitary gates which are involved in the circuit and these issue is still under investigation. In the AdS/CFT case the reference state is clearly \emph{not} the CFT vacuum since the holographic (volume as well as action) for complexity is nonzero for pure AdS dual to the CFT vacuum state. We are unable to shed any further light on this issue here either and as a result our results suffer from the same reference state ambiguity. However since the LST$_2$ under consideration is obtained as a irrelevant deformation of a CFT$_2$ we can imagine using the \emph{same exact} unitary gates and the \emph{same exact} reference state as used for the initial CFT$_2$ which we UV-deformed. This is reasonable since the complexity expressions obtained here reduce to the familiar pure AdS expression once the UV deformation is removed. We \emph{can} say something about the target state though. In the CFT$_2$ case the target state of the zero temperature geometry was the CFT vacuum, invariant under the $SL(2,\mathbb{R})\times SL(2,\mathbb{R})$ symmetry. In the LST$_2$ case, the target state is the ``no string" vacuum
state which is the vacuum of the BRST cohomology of the coset $\frac{SL(2,\mathbb{R}) \times U(1)}{U(1)}$ at zero temperature \cite{Chakraborty:2020yka}. For the finite temperature the target is the thermofield double state, both in case of the CFT and the LST, defined by $\big|\psi \rangle = \sum_n e^{-\beta \omega_n/2} \big|n\rangle_1 \otimes \big| n\rangle_2$ for energy eigenstates $\big|n\rangle$.

For interesting works on complexity in the context of double trace $T\bar{T}$ deformed CFT see \cite{Akhavan:2018wla,Jafari:2019qns,Geng:2019yxo}.

\section{Review of string theory in $AdS_3$, single trace $T\bar{T}$ and LST}\label{sec2}
 
 Let us consider critical superstring background $AdS_3\times \mathcal{M}$ that preserves $\mathcal{N}=2$ or more supersymmetry where $\mathcal{M}$ is a compact spacelike manifold of dimension seven. A well studied example of this kind is type II strings on  $AdS_3\times S^3\times T^4$ that preserves $(4,4)$ supersymmetry. The worldsheet theory describing strings propagating in $AdS_3$ with NS-NS fluxes turned on and R-R fluxes switched off is described by the WZW sigma model on the group manifold $SL(2,\mathbb{R})$. The worldsheet theory is invariant under the left and right moving component of $sl(2,\mathbb{R})$ current algebra at level $k$. The radius of $AdS_3$, $R_{ads}$, is related to the level of the current algebra as $R_{ads}=\sqrt{k}l_s$, where $l_s=\sqrt{\alpha'}$ is the string length.
 
 Via the AdS/CFT correspondence, string theory on $AdS_3$ is dual to a two-dimensional CFT living on the boundary of $AdS_3$. For supergravity approximation to be reliable, we will consider $k\gg1$. In the presence of the NS-NS three form H-flux, the spacetime theory has the following properties:
 \begin{enumerate}
\item{The spacetime theory has a normalizable $SL(2,\mathbb{C})$ invariant vacuum:\begin{itemize}
\item{The NS vacuum, which corresponds to global $AdS_3$ in the bulk.}
\item{The R vacuum, that corresponds to massless ($M=J=0$) BTZ in the bulk.}
\end{itemize}}
\item{The NS sector states contain a sequence of discrete states coming from the discrete series representation of $SL(2,\mathbb{R})$ followed by a continuum of long strings. The continuum starts above a gap of order $\frac{k}{2}$ \cite{Maldacena:2000hw}.}
\item{The R-sector states contain a continuum above a gap of order $\frac{1}{k}$. Here the status of the discrete series states is not quite clear.}
\end{enumerate}
In the discussion that follows, we will focus only the long strings in the R-sector.
   
It was argued in \cite{Seiberg:1999xz} that, for string theory on $AdS_3\times\mathcal{M}$, the theory living on a single long string is described by a sigma model on 
\begin{eqnarray}\label{Mlong}
\mathcal{M}^{(L)}_{6k}=\mathbb{R}_\phi\times \mathcal{M}~,
\end{eqnarray}
with central charge $6k$. The theory on $\mathbb{R}_\phi$ has a dilaton that is linear in $\phi$ with a slope given by
\begin{eqnarray}\label{Qlong}
Q^{(L)}=(k-1)\sqrt{\frac{2}{k}}~.
\end{eqnarray}
The theory on the long strings has an effective coupling given by $\exp({Q^{(L)}\phi})$. Thus the dynamics of the long strings becomes strongly coupled as they move towards the boundary. But there is a wide range of positions on the radial direction where the long strings are weakly coupled.  A natural question that one may ask at this point is: what is the full boundary theory dual to string theory in $AdS_3$. The answer to that question is, in general (for generic $k$), not known, but there are evidences to believe that the theory on the long strings are well described by the symmetric product CFT 
\begin{eqnarray}\label{longsym}
(\mathcal{M}^{(L)}_{6k})^p/S_p~,
\end{eqnarray}
where $p$ can be thought of as the number of fundamental (F1) strings that form the background.
   
String theory in $AdS_3$ contains an operator $D(x,\bar{x})$ \cite{Kutasov:1999xu} (where $x$ and $\bar{x}$ are coordinates of the two-dimensional spacetime theory), in the long string sector that has many properties in common with the $T\bar{T}$ operator. For example $D(x,\bar{x})$ is a $(2,2)$ quasi-primary operator of the spacetime Virasoro and has the same OPE with the stress tensor as the $T\bar{T}$ operator. However, there is an important difference between the $T\bar{T}$ operator and the operator $D(x,\bar{x})$: 
 $T\bar{T}$ is a double trace operator whereas $D(x,\bar{x})$ is single trace.\footnote{Here single trace refers to the fact that $D(x,\bar{x})$ can be expressed as a single integral over the worldsheet of a certain worldsheet vertex operator. The operator $T\bar{T}$ on the other hand is double trace because it can be expressed as a product of two single trace operators in the sense just described.} In fact 
\begin{eqnarray}\label{D}
D(x,\bar{x})=\sum_{i=1}^pT_i\bar{T}_i~,
\end{eqnarray}
where $T_i\bar{T}_i$ can be thought of as the $T\bar{T}$ operator of the $i^{th}$ block  $\mathcal{M}^{(L)}_{6k}$ in the symmetric product CFT $(\mathcal{M}^{(L)}_{6k})^p/S_p$. For an elaborate discussion along this line see \cite{Chakraborty:2019mdf,Chakraborty:2018vja}
   
Next, let us consider deformation of the long string symmetric product by the operator $D(x,\bar{x})$. This corresponds to deforming the $i^{th}$ block CFT $\mathcal{M}^{(L)}_{6k}$ by the operator $T_i\bar{T}_i$ and then symmetrized. Note that such a deformation is irrelevant and it involves flowing up the renormalization group (RG) trajectory. The deformation of the spacetime theory by  $D(x,\bar{x})$ induces on the worldsheet a truly marginal deformation:
 \begin{eqnarray}\label{wsind}
 \int_{(\mathcal{M}^{(L)}_{6k})^p/S_p} d^2 xD(x,\bar{x})\sim\int_{\Sigma} d^2z J_{SL}^-\bar{J}_{SL}^-~,
 \end{eqnarray}
 where $z,\bar{z}$ are the complex coordinates of the worldsheet Riemann surface $\Sigma$, $J_{SL}^-$ and $\bar{J}_{SL}^-$ are respectively the left and right moving null $sl(2,\mathbb{R})$ currents of the worldsheet theory.     
   
The above current-anti-current deformation of the worldsheet $\sigma-$model is exactly solvable, and standard worldsheet techniques yield the metric (in string frame), dilaton and the B-field as \cite{Forste:1994wp,Israel:2003ry}
 \begin{eqnarray}
 \begin{split}\label{M3}
& ds^2=f^{-1}(-dt^2+dx^2)+kl_s^2\frac{dU^2}{U^2}~,\\
& e^{2\Phi}=\frac{g_s^2}{kU^2}f^{-1}~,\\
& dB=\frac{2i}{k^{3/2}\,l_s\,U^2}f^{-1}\epsilon_3~,
  \end{split}
 \end{eqnarray}
 where $f=\lambda+\frac{1}{kU^2}$, $\lambda$ is the dimensionless coupling \footnote{Note that without loss of generality, the value of $\lambda$ can be set to an appropriate value as discussed in \cite{Giveon:2017nie}.} of the marginal worldsheet deformation  and $g_s$ is the asymptotic string coupling in $AdS_3$ with $g_s^2=e^{2\Phi(U\to 0)}\equiv e^{2\Phi_0}$. This background is popularly known as $\mathcal{M}_3$. The background $\mathcal{M}_3$ \eqref{M3} interpolates between $AdS_3$ in the IR (\ie\ $U\ll 1/\sqrt{k\lambda}$) to flat spacetime with a linear dilaton, $\mathbb{R}^{1,1}\times\mathbb{R}_\phi$ in the UV (\ie\ $U\gg 1/\sqrt{k\lambda}$). The coupling $\lambda$ sets the scale at which the transition happens. 
 
 The deformed sigma model background \eqref{M3} can also be obtained as a solution to the equations of the motion of  three dimensional supergravity action \cite{Giveon:2005mi,Chakraborty:2020yka}
 \begin{eqnarray}\label{sugra}
 S= \frac{1}{16\pi G_N}\int d^3 X \sqrt{-g}e^{-2(\Phi-\Phi_0)}\left( R+4g^{\mu\nu}\partial_\mu\Phi\partial_\nu\Phi-\frac{1}{12}H^2-4\Lambda\right)~,
 \end{eqnarray} 
where $G_N$ is the three-dimensional Newton's constant in $AdS_3$, $g_{\mu\nu}$ is the string frame metric, $R$ is the Ricci scalar (in string frame), $\Phi$ is the dilaton, $H=dB$ is the 3-form flux and $\Lambda$ is the cosmological constant.

As an example, the above construction can be realized as follows. Let us consider a stack of $k$ NS5 branes in flat space wrapping a four dimensional compact manifold (\eg\ $T^4$ or $K_3$). The near horizon geometry of the stack of $k$ NS5 branes is given by $\mathbb{R}^{1,1}\times\mathbb{R}_\phi$ with a dilaton that is linear in the radial coordinate $\phi$ (where $\phi=\log(\sqrt{k} U)$). The string coupling goes to zero near the boundary (\ie\ $U\to \infty$) whereas it grows unboundedly as one goes deep in the bulk (\ie\ $U\to 0$).  Next, let's add $p$ (with $p\gg1$) F1 strings stretched along $\mathbb{R}^{1,1}$. This stabilizes the dilaton and the string coupling saturates as $g_s\sim 1/\sqrt{p}$. Thus for large $p$ the string coupling is weak and one can trust string perturbation theory. The F1 strings modifies the IR geometry (\ie\ $U\ll 1/\sqrt{k\lambda}$) to $AdS_3$. The smooth interpolation between   $\mathbb{R}^{1,1}\times\mathbb{R}_\phi$ in the UV to $AdS_3$ in the IR corresponds to interpolation between near horizon geometry of the NS5 brane system to that of the F1 strings \cite{Chakraborty:2020swe,Chakraborty:2020yka}. The spacetime theory interpolates between a CFT$_2$ with central charge $6kp$ in the IR to two-dimensional LST in the UV. The theory is nonlocal in the sense that the short distance physics is not governed by a fixed point.

 LST can be realized as the decoupled theory on the NS5 branes.  
It has properties that are somewhat intermediate between a local quantum field theory and a full fledged critical string theory. Unlike a local field theory, at high energy $E$, LST has a Hagedorn density of states $\rho\sim e^{\beta_HE}$ where $\beta_H=2\pi l_s\sqrt{k\lambda }$. On the other hand, LST has well defined off-shell amplitudes \cite{Aharony:2004xn} and upon quantization it doesn't give rise to massless spin 2 excitation. Both these properties are very similar to  local quantum field theories. For a detailed review of LST see \cite{Aharony:1998ub,Kutasov:2001uf}

The above discussion has a simple generalization to backgrounds at finite temperature \cite{Giveon:2017nie,Chakraborty:2020swe,Apolo:2019zai}:
\begin{eqnarray}
 \begin{split}\label{M3T}
& ds^2=-\frac{f_1}{f}dt^2+\frac{1}{f}dx^2+kl_s^2\,{f_1}^{-1}\frac{dU^2}{U^2}~,\\
& e^{2\Phi}=\frac{g_s^2}{kU^2}f^{-1}~,\\
& dB=\frac{2i}{k^{3/2}\,l_s\,U^2}f^{-1}\epsilon_3~,
  \end{split}
 \end{eqnarray}
 where   as before $f=\lambda+\frac{1}{kU^2}$ and $f_1=1-\frac{U_T^2}{U^2}$ where $U_T$ is the radius of the outer horizon of the black hole. There is also an inner horizon at $U=0$. For the Penrose diagram see figure \ref{fig:finiteT M3 WdW}. From the worldsheet sigma model point of view, the above background can be obtained from the coset description $\frac{SL(2,\mathbb{R})\times U(1)}{U(1)}$ \cite{Giveon:2005mi,Chakraborty:2020yka,Apolo:2019zai}. One can also check that solution \eqref{M3T} satisfies the equations of motion obtained from the supergravity action \eqref{sugra}.
 
 Going to the Euclidean continuation, and demanding the smoothness of the metric at the horizon, one can read off the temperature of the black hole \eqref{M3T} as
 \begin{eqnarray}\label{bhtemp}
 T_{bh}=\frac{1}{2\pi l_s}\frac{U_T}{\sqrt{1+\lambda k U_T^2}}~.
 \end{eqnarray}
   When $U_T\ll \frac{1}{\sqrt{k\lambda}}$, the horizon sits deep inside the bulk where the local geometry is well approximated by $AdS_3$. To good approximation such a black hole is described by BTZ. For $U_T\gg \frac{1}{\sqrt{k\lambda}}$ the horizon sits in the asymptotic linear dilaton regime of the geometry. The black hole here is well described by coset $\frac{SL(2,\mathbb{R})}{U(1)}\times U(1)$. 
   
   As $U_T$ increases the black hole temperature \eqref{bhtemp} increases but saturates to an Hagedorn temperature 
   \begin{eqnarray}
   \beta_H=\frac{1}{T_H}=2\pi l_s\sqrt{k\lambda}~,
   \end{eqnarray}
   as $U_T\to\infty$. This is an indication of the Hagedorn nature of the spacetime theory (LST) in the UV.
   
   Note that in the discussion that follows, we will consider only the positive sign of the coupling $\lambda$. In that case the spectrum of the deformed theory is real and the theory is unitary. Holography in the background \eqref{M3} and \eqref{M3T} has been studied extensively in \cite{Giveon:2017nie,Giveon:2017myj,Asrat:2017tzd,Giribet:2017imm,Chakraborty:2018kpr,Chakraborty:2018aji,Chakraborty:2020yka}. For the other sign of the coupling see \cite{Chakraborty:2020swe,Chakraborty:2020cgo,Aguilera-Damia:2020qzw}.

\section{Holographic Complexity in $\mathcal{M}_3$ at zero temperature}   \label{0T complexity}
 
The aim of this section to compute the computational complexity of the LST dual to the background $\mathcal{M}_3$ \eqref{M3} using  holographic methods, namely the Complexity-Volume (CV) \cite{Susskind:2014rva} and Complexity-Action (CA) \cite{Brown:2015bva, Brown:2015lvg} prescriptions. We will perform these complexity computations for \emph{both} zero temperature (in section \ref{0T complexity}) and finite temperature cases (in section \ref{finiteT}). Computational complexity like entanglement entropy, is a manifestly UV-divergent quantity, and for local quantum field theories the UV divergence structure of computational complexity is rigidly constrained \cite{ Carmi:2016wjl, Reynolds:2016rvl}. In this section we reveal the UV-divergences which arise in a nonlocal field theory such as two-dimensional LST, and compare and contrast them with those arising in a local quantum field theory (\eg\ a CFT$_2$).

\subsection{Volume complexity at zero temperature}

The volume complexity prescription computes the complexity of the dual boundary theory in terms of the volume of a maximal volume spacelike slice , $\Sigma$,
\begin{equation}
C_V = \frac{V_{\Sigma}}{G_N\,L}~,\ \  \text{ with } \ \   V_\Sigma = \int_\Sigma d^{D-1}x\, \sqrt{\gamma_\Sigma}~,
\end{equation}
where $\gamma_{\mu\nu}$ is the pullback metric on the maximal volume slice. As mentioned before, $L$ represents a suitable characteristic scale of the geometry. However, we are working in the string frame with a non-trivial dilaton background and the volume complexity proposal needs to be generalized. The appropriate generalization is given by,
\begin{equation}
C_V  = \frac{ \tilde{V}_{\Sigma}}{\kappa_0^2\, L}~, \ \ \ \text{with } \ \  \tilde{V}_{\Sigma} = \int_\Sigma d^{D-1}x\,  e^{-2\left(\Phi-\Phi_\infty\right)} \sqrt{\gamma_\Sigma}~. \label{CV_str}
\end{equation}
One can check that this generalization furnishes the correct powers of $G_N$ \footnote{See \cite{Klebanov:2007ws} for a similar prescription for the Ryu-Takayanagi formula for the entanglement entropy} in the denominator using the string convention, $\kappa_0^2 e^{-2(\Phi_\infty-\Phi_0)} = 8 \pi G_N$ where $e^{\Phi_\infty}$ is the flat space string coupling and $e^{\Phi_0}$ is the string coupling of $AdS_3$.

For the putative (string frame) maximal volume spacelike surface $\Sigma$ given by  $t=t(U)$, in the zero temperature $\mathcal{M}_3$ geometry \eqref{M3},  the induced metric  is
\begin{eqnarray}
ds_\Sigma^2 \equiv \gamma_{ab} dx^a dx^b=\left(\frac{k\,l_s^2}{U^2}-f^{-1} t'(U)^2\right)dU^2+f^{-1}\,dx^2~, \ \ \text{ where } \ \ t'\equiv \frac{dt}{dU}~.
\end{eqnarray}
In the string frame, the volume of such a spacelike slice anchored at a time $T$ \footnote{The $T$ here is not to be confused with the temperature $T_{bh}$ \eqref{bhtemp} in section \ref{finiteT}.\label{f7}} on the boundary is,
\begin{equation}
\begin{split}
\tilde{V}(T) &= e^{2(\Phi_\infty-\Phi_0)} \int dx\, dU \,  e^{-2\left(\Phi-\Phi_0 \right)} \sqrt{\gamma_{\Sigma}} \\
&= \frac{ k^{3/2} l_s L_x}{e^{-2(\Phi_\infty-\Phi_0)}} \int_0^{\infty} dU\,U\, f^{{1/2}}\,\sqrt{1-\frac{U^2\,t'(U)^2}{k\,l_s^2\,f} }~ .
\end{split}
\end{equation} 
Here $L_x=\int dx$ is the spatial extent (IR cutoff) of the boundary theory target space.  Extremizing this volume leads to the following Euler-Lagrange equation:
\begin{eqnarray}
U \left(1+\lambda \,k\,U^2\right)t''+\left(4+3\lambda \,k\,U^2\right)t'  -\frac{2U^4}{l_s^2} {t'}^3 =0~. \label{EL_0T}
\end{eqnarray}
The solution is found by employing series expansion method, lets assume the near boundary expansion of $t(U)$ of the form:
\begin{eqnarray}\label{maxvslice}
t(U ) & = & T+\frac{a_1}{U }+ \frac{a_2}{U^2}+\frac{a_3}{U ^3}+\dots ~.
\end{eqnarray}
And plugging back in \eqref{EL_0T} and solving them order by  order in $\frac{1}{U}$, we obtain the result that all the coefficients vanish. Thus the maximal volume slice is $t(U)=T$, a result that can be anticipated from the time reflection symmetry: $t\rightarrow-t$, of the background. Thus, the volume of the maximal volume slice is,
\begin{equation}
\tilde{V}_\Sigma(T)=\frac{ k^{3/2}\, l_s\, L_x}{e^{-2(\Phi_\infty-\Phi_0)}} \int_0^{\infty} dU\,U\, f^{{1/2}}=\frac{ k\, l_s\, L_x}{e^{-2(\Phi_\infty-\Phi_0)}}\int_0^\infty dU \sqrt{1+k\lambda U^2}~, 
\end{equation}
which diverges as $U \rightarrow \infty$. So we impose a UV cutoff at $U=l_s/\epsilon$ to regulate it. The regulated volume is then,
\begin{eqnarray}
\tilde{V}_\Sigma(T)  = \frac{k\,l_s\,L_x}{e^{-2(\Phi_\infty-\Phi_0)}}\left[\frac{l_s}{2\epsilon} \sqrt{1+\frac{k\,\lambda\,l_s^2}{\epsilon^2}}  +\frac{\sinh^{-1}\left(\frac{\sqrt{k\,\lambda}\,l_s}{\epsilon}\right)}{2\sqrt{k\,\lambda}} \right]~.
\end{eqnarray}
As expected, due to time translation symmetry the expression is independent of $T$. Therefore from \eqref{CV_str} volume complexity turns out to be:
\begin{eqnarray}
C_{V} \equiv  \frac{\tilde{V}_\Sigma}{\kappa_0^2\,L}  & = & \frac{k\,l_s\,L_x}{G_N\,L}\left[\frac{l_s}{2\epsilon} \sqrt{1+\frac{k\,\lambda\,l_s^2}{\epsilon^2}}  +\frac{\sinh^{-1}\left(\frac{\sqrt{k\,\lambda}\,l_s}{\epsilon}\right)}{2\sqrt{k\,\lambda}} \right]~.
\end{eqnarray}
Note that by convention  the length scale $L$ appearing here is the characteristic length scale associated with the geometry. Comparison with results from action complexity  helps us resolve this ambiguity $L=\ell=\sqrt{k}\,l_s$, the AdS radius, and the volume complexity is thus,
\begin{eqnarray}
C_V=\frac{cL_x}{3\beta_H}\left[\frac{\beta_H}{2\epsilon}\sqrt{4+\frac{\beta_H^2}{\pi^2\epsilon^2}}+2\pi~\sinh^{-1}\left(\frac{\beta_H}{2\pi\epsilon}\right)\right]~,\label{CVl}
\end{eqnarray}
where $c$ is the Brown-Henneaux central charge of the undeformed CFT$_2$  given by
\begin{eqnarray}
c=\frac{3\sqrt{k}l_s}{2G_N}~.\label{BHcc}
\end{eqnarray}

\subsubsection{A comment on the non-locality: An \emph{``effective central charge"} for LST}\label{lstlim}

Let us recall that $\beta_H$  can be thought of the length scale below which non-locality kicks in. Thus, an  interesting limits to study would be $\epsilon/\beta_H\ll1$ where the short distance physics is that of a non-local theory. In this limit the volume complexity takes the form
\begin{eqnarray}
\lim_{\epsilon/\beta_H\to 0} C_{V} &=&\frac{cL_x}{3\beta_H}\left[\frac{\beta_H^2}{2\pi \epsilon^2}+2\pi \log\left(\frac{\beta_H}{\pi\epsilon}\right)+ \pi +O\left(\frac{\epsilon}{\beta_H}\right)\right]~.\label{CV@0T}
\end{eqnarray}
Evidently the divergence structure of the volume complexity \eqref{CV@0T} does not appear like that of a local quantum field theory. 

For the case of a local quantum field theory, complexity being an extensive quantity should be proportional to the degrees of freedom given by the number of lattice sites $\propto L_x/\epsilon$ \ie\ scales inversely with the cutoff $\epsilon$ (lattice spacing). 
The quadratic and logarithmic divergences in \eqref{CV@0T} are a reflection of the fact that the boundary theory, being a LST, is a non-local field theory and fittingly the non-locality parameter $\beta_H$ features in the coefficient of this quadratic as well as the logarithmic divergences. One can check by making the non-locality vanish \ie\ in the limit $\epsilon/\beta_H\gg1$, the volume complexity expression \eqref{CVl} indeed reduces to that of a local field theory,
\begin{equation}\label{CVir}
\lim_{\epsilon/\beta_H\gg 1 } C_V =\frac{2c}{3\beta_H}~\frac{L_x}{(\epsilon/\beta_H)}= \frac{2c}{3}~\frac{L_x}{\epsilon}~. 
\end{equation}
This expression of complexity (being proportional to the product of $c$, the central charge \ie\ the number of degrees of freedom per lattice site, and $L_x /\epsilon$, which gives the total number of lattice sites) counts the total number of degrees of freedom  in a local field theory.

Now a remarkable physical fact emerges when one considers the coefficient of the log term (which is universal) in the expression of volume complexity \eqref{CV@0T} in the deep UV (\ie\ $\epsilon\ll \beta_H$), which is 
 \begin{equation}
\tilde{N}= c\frac{L_x}{\beta_H}~. \label{total dof}
\end{equation}
This coefficient counts the total number of  ``regularized/effective" degrees of freedom in the theory if we regard the lattice spacing of LST to be the Hagedorn scale, $\beta_H$ instead of the UV cutoff $\epsilon$ of the original IR CFT, namely, $ c \frac{L_x}{\epsilon}$. 

Another interesting fact emerges when we focus on the quadratic divergence in \eqref{CV@0T}. One can rewrite this term in a manner which ``looks" like a local field theory as follows,
\begin{equation}
C_V = \, \frac{c L_x \beta_H}{6\pi ^2 \epsilon^2} + \ldots =\,  \frac{2\tilde{c}(\epsilon)}{3} \frac{L_x}{\epsilon} + \dots, \ \text{ where }\ \ \tilde{c}(\epsilon)= c \frac{\beta_H}{4 \pi^2 \epsilon}~,
\end{equation}
where $\tilde{c}(\epsilon)$ now has to be interpreted as an \emph{``effective central charge''} for LST which is a monotonically increasing function of UV energy scale, $\frac{1}{\epsilon}$, and in particular this ``effective central charge" diverges as the UV cutoff is removed. 

\begin{figure}[h]
    \centering
    \includegraphics[width=.5\textwidth]{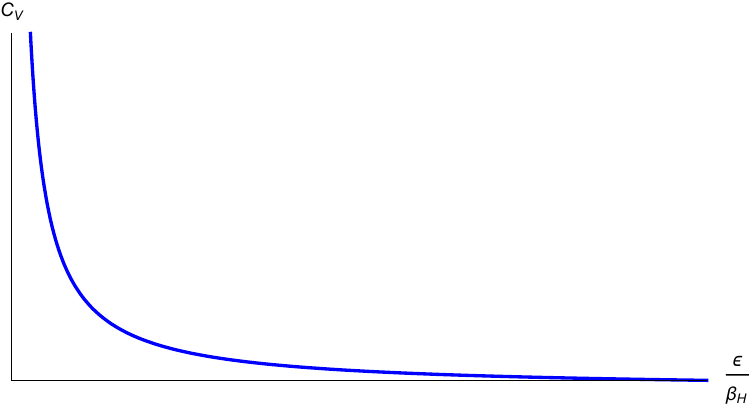}
    \caption{$C_V(\epsilon/\beta_H)$ vs $\epsilon/\beta_H$ at $T=0$.}
    \label{fig1}
\end{figure}

The full volume complexity \eqref{CVl} as a function of $\epsilon/\beta_H$ has the following interesting properties:
\begin{enumerate}
\item{$C_V$ in \eqref{CVl} as a function of $\epsilon/\beta_H$ is always positive and monotonically decreases from UV to IR (\ie\ $C_V'(\epsilon/\beta_H)\geq 0$).}
\item{In the deep UV (\ie\ for $\epsilon/\beta_H\ll 1$), $C_V$ diverges as \eqref{CV@0T}.}
\item{In the deep IR (\ie\ for $\epsilon/\beta_H\gg 1$), $C_V$ decreases to 0 as \eqref{CVir}.}
\end{enumerate}
The complete variation $C_V$ as a function of $\epsilon/\beta_H$ is given in figure \ref{fig1}.

\subsection{Action complexity  at zero temperature}
\begin{figure}[h]
 \centering
 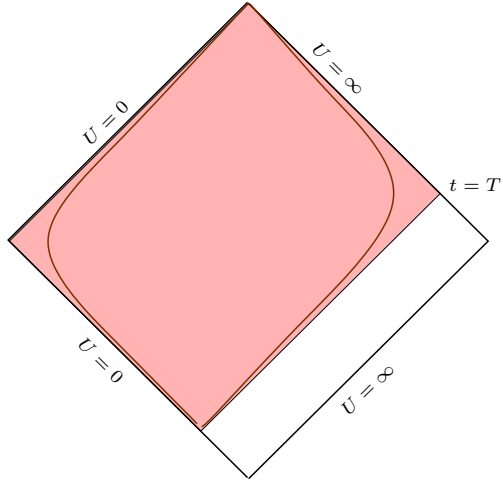
  \caption{Penrose diagram of the $\mathcal{M}_3$ geometry with the Wheeler-deWitt (WdW) patch shaded in pink for the boundary time $T$. The brown curves are timelike surfaces which can be continuously deformed into the null boundaries of the WdW patch.}
  \label{fig:0T M3 WdW}
\end{figure}

Now we compute the action complexity, $C_\mathcal{A}$, for the zero temperature $\mathcal{M}_3$ geometry. Action complexity has the dual advantage that (a) there are no arbitrary length scales appearing in its definition, and (b) neither does one need to solve a variational problem (maximal volume). Instead one just performs action integrals over the so called \emph{WdW patch} which is defined to be the union of all spacelike curves in the bulk anchored at a fixed time slice on the boundary:
\begin{equation}\label{caa}
C_\mathcal{A}=\frac{S_{WdW}}{\pi \hbar}~.
\end{equation}
The Penrose diagram of the $\mathcal{M}_3$ spacetime with the WdW patch is displayed in figure \ref{fig:0T M3 WdW}.

The gravity action in the string frame is:
\begin{eqnarray}
\begin{split}\label{fullaction}
S &= \frac{1}{16\pi G_N}\int_{M} d^3 X \sqrt{-g}e^{-2(\Phi-\Phi_0)}\left( R+4g^{\mu\nu}\partial_\mu\Phi\partial_\nu\Phi-\frac{H^2}{12}-4\Lambda\right) \\
&   \hspace{1in}+ \frac{1}{8\pi\,G_N} \int_{\sum \partial M} \sqrt{\gamma}\, \left( \cdots\right) + \frac{1}{8\pi\,G_N} \int_{\cap \partial M} \sqrt{h}\, (\cdots)~.
\end{split}
\end{eqnarray}
The $(\cdots)$'s represent the appropriate surface/boundary ($\cup \partial M$) terms  and joint ($\cap \partial M$) terms needed to make the variation of the action well defined as well as reparametrization invariant.  Since (some) boundaries of the WdW patch are null, the usual GHY terms are not the suitable ones. This issue of determining the boundary terms for null boundaries was settled in \cite{Lehner:2016vdi}. However, we will take an alternative prescription spelled out in \cite{Bolognesi:2018ion}\footnote{see also \cite{Parattu:2015gga}.} where the null boundaries of the WdW patch are first deformed into a single smooth timelike surface using a deformation parameter (regulator), and then we are free to use the usual GHY term. After working out the GHY term we remove the regulator and obtain the result for the null WdW boundary. This affords an enormous simplification as it eliminates the necessity to compute the joint terms (\ie\ terms in the action from joints or edges along which two null surfaces intersect) as well as preserving diffeomorphism and reparametrization invariance of the GHY contribution from beginning to end. Our regularization reproduces the same results as the prescription of \cite{Lehner:2016vdi} for the well known cases of pure AdS, AdS-Schwarzschild, AdS-RN etc. but the status of the equivalence of these two prescriptions for arbitrary generic geometries is yet unexplored. In general the issue of different regularization prescriptions is still being investigated e.g. for a comparison of the two regularizations introduced in \cite{Carmi:2016wjl}, see \cite{Akhavan:2018wla, Jafari:2019qns}.

There is an additional issue regarding boundary terms here since we are working in the string frame while the usual GHY term applies for surface terms in the Einstein frame. In the string frame the GHY surface term has a contribution from the dilaton factor. This is determined by starting out with the usual GHY term in the Einstein frame and then Weyl transforming the expression to string frame. For $2+1$-dimensional bulk  \footnote{For $D$-dimensions
\begin{equation*}
S_{GHY}=2\int d^2x \sqrt{-h} e^{-2\Phi} \left( K- 2\left(\frac{D-1}{D-2}\right)\, n^M\partial_M\Phi\right).
\end{equation*}}, the string frame GHY term is
\begin{equation}
S_{GHY}=\frac{1}{8\pi G_N}\int d^2x \sqrt{-\gamma}\, e^{-2(\Phi-\Phi_0)} \left( K- 4n^M\partial_M\Phi\right)~. \label{string frame GHY}
\end{equation}

\subsubsection{Volume (EH) pieces of the onshell action}
The volume terms in the bulk action \eqref{fullaction} are
\begin{eqnarray}
S=\frac{1}{16\pi\,G_N} \int_{WdW} d^3 X \sqrt{-g}e^{-2(\Phi-\Phi_0)}\left( R+4g^{\mu\nu}\partial_\mu\Phi\partial_\nu\Phi-\frac{1}{12}H^2-4\Lambda\right). \label{bulk action}
\end{eqnarray}
For the zero temperature $\mathcal{M}_3$ background, the Ricci scalar is,
\begin{equation}
R=\frac{-6+8\lambda\,k\,U^2}{k\,l_s^2\,\left(1+\lambda\,k\,U^2\right)^2}~,
\end{equation} and the dilaton is given by,
\begin{equation}
\Phi = \Phi_{0} -\frac{1}{2}\ln\left(1+\lambda\,k\,U^2\right)~.
\end{equation}
The Wheeler-deWitt patch (WdW) for the boundary time $t=T$ is bounded by the null rays
\begin{eqnarray}
\,dt_{\pm}&=& \mp \sqrt{k}\,l_s\,\frac{\sqrt{f}}{U}dU~, 
\end{eqnarray}
obeying boundary condition, $t(U \rightarrow\infty)=T$. The $t$-integrals in the volume terms \eqref{bulk action} (Einstein-Hilbert terms) can be readily done:

\begin{equation}
t_{+}(U)-t_{-}(U) =2\sqrt{k}\,l_s\,\int_{U}^{\infty}\, dU' \, \frac{\sqrt{f(U')}}{U'}~.
\end{equation}
This integral is divergent and hence we will modify our WdW patch to begin at a UV-cutoff surface $U=l_s/\epsilon$ instead of spatial infinity:
\begin{equation}
t_{+}(U)-t_{-}(U) =2\sqrt{k}\,l_s\,\int_{U}^{l_s/\epsilon}\, dU' \  \frac{\sqrt{f(U')}}{U'}~.
\end{equation}
Various bulk contributions are listed as follows (in the intermediate steps one may consider the change of variables  $U\rightarrow z = \frac{U}{l_s/\epsilon}$ and $U'\rightarrow z'  = \frac{U'}{l_s/\epsilon}$ to perform the integrals exactly).\\

\noindent \textbf{The Ricci scalar term in the action:}
\begin{eqnarray}
\begin{split}\label{s1}
S_{R}&\equiv  \frac{1}{16\pi G_N}\int_{WdW} d^3 X \sqrt{-g} ~ e^{-2(\Phi-\Phi_0)}\, R \\
& =  \frac{k L_x}{8 \pi G_N} \int_0^{l_s/\epsilon} dU\,U\, \frac{-6 + 8 \lambda k U^2}{\left(1+\lambda k U^2\right)^2} \int_U^{l_s/\epsilon} \frac{dU'}{U'} \sqrt{f(U')}~.
\end{split}
\end{eqnarray}
The above integral can be performed analytically but the full expression is a bit cumbersome. In the deep UV (\ie\ when $\epsilon/\beta_H\ll1$), $S_R$ takes the following form
\begin{eqnarray}
\begin{split} \label{SR}
\lim_{\epsilon/\beta_H\ll 1}S_R=&-\frac{cL_x}{6\beta_H}(7+8\log2)\log\left(\frac{\beta_H}{\pi\epsilon}\right)+\frac{2cL_x}{3\beta_H}\log^2\left(\frac{\beta_H}{\pi\epsilon}\right) \\
&+\frac{cL_x}{18\beta_H}(\pi^2+24\log2)+O\left(\epsilon^2/\beta_H^2\right) ~.
\end{split}
\end{eqnarray}
In the IR (\ie\ when $\epsilon/\beta_H\gg1$), $S_R$ takes the form
\begin{eqnarray}
\lim_{\epsilon/\beta_H\gg 1}S_R= -\frac{cL_x}{4\pi\beta_H}\frac{\beta_H}{\epsilon}+\frac{7cL_x}{288\pi^3\beta_H}\left(\frac{\beta_H}{\epsilon}\right)^3+O\left(\beta_H^4/\epsilon^4\right)~.
\end{eqnarray}

\noindent \textbf{The dilaton kinetic term in the action:}
\begin{eqnarray}
\begin{split}\label{s2}
S_{\Phi}&\equiv& \frac{1}{16\pi G_N}\int_{WdW} d^3X \sqrt{-g}\, e^{-2(\Phi-\Phi_0)} \left( 4g^{\mu \nu } \partial_\mu \Phi \partial_\nu \Phi\right)\\
&=& \frac{L_x k^3 \lambda^2}{2 \pi G_N} \int_0^{l_s/\epsilon} dU \frac{U^5}{\left(1+\lambda\,k\,U^2\right)^2} \int_U^{l_s/\epsilon} \frac{dU'}{U'}\sqrt{f(U')}~.
\end{split}
\end{eqnarray}
In the UV $S_\Phi$ takes the following form:
\begin{eqnarray}
\begin{split}\label{SPhi}
\lim_{\epsilon/\beta_H\ll 1}S_\Phi=&\frac{cL_x}{24 \pi^2\beta_H}\left(\frac{\beta_H}{\epsilon}\right)^2+\frac{cL_x}{6\beta_H}(3+8\log 2)\log\left(\frac{\beta_H}{\pi\epsilon}\right)-\frac{2cL_x}{3\beta_H}\log^2\left(\frac{\beta_H}{\pi\epsilon}\right)\\
&-\frac{cL_x}{36\beta_H}(-3+2\pi^2+48\log 2)+O\left(\epsilon^2/\beta_H^2\right) ~.
\end{split}
\end{eqnarray}
One might be a bit alarmed at the appearance of the ``log squared" divergences in the expressions \eqref{SR} and \eqref{SPhi}, which did not arise in the volume complexity cases but as it will turns out,  such log squared divergent contributions will cancel out among each other. 

In the IR, $S_\Phi$ takes the form
\begin{eqnarray}
\lim_{\epsilon/\beta_H\gg 1}S_\Phi=0+O\left(\beta_H^5/\epsilon^5\right)~.
\end{eqnarray}

\noindent  \textbf{The cosmological constant term in the action:}
\begin{eqnarray}
\begin{split}\label{s3}
S_{\Lambda}&\equiv  \frac{1}{16\pi G_N}\int_{WdW} d^3 X \sqrt{-g}\,e^{-2(\Phi-\Phi_0)}\, \left(- 4\,\Lambda\right)\\
&=  \frac{L_x k}{2\pi G_N} \int_0^{l_s/\epsilon} dU\,U\int_u^{l_s/\epsilon}\frac{dU'}{U'}\sqrt{f(U')}~.
\end{split}
\end{eqnarray}
In the UV $S_\Lambda$ takes the following form
\begin{eqnarray}\label{Scc}
\lim_{\epsilon/\beta_H\ll 1}S_\Lambda=\frac{cL_x}{24 \pi^2\beta_H}\left(\frac{\beta_H}{\epsilon}\right)^2+\frac{cL_x}{6\beta_H}\log\left(\frac{\beta_H}{\pi\epsilon}\right)+\frac{cL_x}{12\beta_H}+O\left(\epsilon^2/\beta_H^2\right) ~.
\end{eqnarray}
In the IR, $S_\Phi$ takes the form
\begin{eqnarray}
\lim_{\epsilon/\beta_H\gg 1}S_\Lambda=\frac{cL_x}{6\pi\beta_H}\frac{\beta_H}{\epsilon}+\frac{cL_x}{144\pi^3\beta_H}\left(\frac{\beta_H}{\epsilon}\right)^3+O\left(\beta_H^4/\epsilon^4\right)~.   
\end{eqnarray}

\noindent  \textbf{The Kalb-Ramond term in the action:}
\begin{eqnarray}
\begin{split}\label{s4}
S_{H}&\equiv  \frac{1}{16\pi G_N}\int_{WdW} d^3 X \sqrt{-g}\,e^{-2(\Phi-\Phi_0)}\, \left(- \frac{H^2}{12}\right) \\
&= -\frac{L_x}{4\pi G_N\,k}  \int_0^{l_s/\epsilon} \frac{dU}{U^3\,f^2}\int_U^{l_s/\epsilon} \frac{dU'}{U'} \sqrt{f(U')} ~.
\end{split} 
\end{eqnarray}
In the UV $S_H$ takes the following form
\begin{eqnarray}\label{SH}
\lim_{\epsilon/\beta_H\ll 1}S_H=-\frac{cL_x}{6\beta_H}\log\left(\frac{\beta_H}{\pi\epsilon}\right)+O\left(\epsilon^2/\beta_H^2\right) ~.
\end{eqnarray}
In the IR, $S_\Phi$ takes the form
\begin{eqnarray}
\lim_{\epsilon/\beta_H\gg 1}S_H=-\frac{cL_x}{12\pi\beta_H}\frac{\beta_H}{\epsilon}+\frac{cL_x}{288\pi^3\beta_H}\left(\frac{\beta_H}{\epsilon}\right)^3+O\left(\beta_H^4/\epsilon^4\right)~. 
\end{eqnarray}

\subsubsection{Surface term at $U=0$}
This is the AdS Poincar\'{e} horizon which is a null surface on which the induced metric $h$ degenerates. Instead we will work with the timelike surface, $U=\delta $, evaluate the GHY term and take the limit, $\delta \rightarrow 0$ of the final expression. The metric on this timelike surface, $U=\delta$, is,
\begin{equation}
ds^2= \frac{1}{f} \left(-dt^2+dx^2\right) ~.
\end{equation} 
The components of the unit outward normal vector for such a constant $U$ surface are:
\begin{equation}\label{normal1}
n^{U}=-\frac{U}{\sqrt{k} l_s}~,   \ \ \ \ \ \ \ \  n^t=n^{x}=0~.
\end{equation}
Using the Christoffel Symbols:
\begin{equation}\label{chris}
\Gamma^{U}_{UU}=-\frac{1}{U}~,   \ \ \ \ \ \ \ \  \Gamma^t_{t\rho}= \Gamma^{x}_{x\,\rho}= -\frac{1}{f}\frac{df}{dU}~,
\end{equation}
  and the unit norma vector \eqref{normal1}, we get the extrinsic curvature of $U=\delta$ surface,
\begin{equation}
K=-\frac{2}{\sqrt{k}\,l_s\,\left(1+\lambda k \delta^2\right)}~.
\end{equation}
The GHY surface term at the Poincar\'e horizon
\begin{eqnarray}
\begin{split}
S^0_{GHY}&=\lim_{\delta \rightarrow0}\,\frac{1}{8\pi G_N} \int dx\int_{t_{-}(\delta)}^{t_{+}(\delta)}dt \,\sqrt{-\gamma(\delta)\,}e^{-2(\Phi-\Phi_0)}\left(K-4n^\rho\partial_\rho \Phi\right) \\
&=\lim_{\delta \rightarrow0} \frac{2\,L_x}{8\pi G_N}\, k\delta^2\,\left(\frac{-2-4 \lambda k \delta^2}{1+\lambda k \delta^2}\right) \int_\delta^{l_s/\epsilon} \frac{dU'}{U} \sqrt{f(U)}=0~.  \label{SP}
\end{split}
\end{eqnarray}

\subsubsection{Action Contributions from the null boundaries of the WdW patch}

The null boundaries of the WdW patch are defined by 
\begin{equation}
(t-T) =\mp \sqrt{k}\,l_s A(U)~; \quad \text{where }\ \ A(U)\equiv \int_{l_s/\epsilon}^U dU' \frac{dU' \sqrt{f(U')}}{U'}~;
\end{equation}
where $T$ is defined in \eqref{maxvslice}.
However, we will deform the pair of null surfaces to a single smooth timelike surface  by introducing a dimensionless parameter, $\varepsilon$,\footnote{This is distinct from the UV regulator, $\epsilon$.}
\begin{equation}
\frac{(t-T)^2}{k\,l_s^2} - (1+\varepsilon) A^2(U) =0~. \label{GHYbdry}
\end{equation}
Taking differentials of both sides leads to,
\begin{eqnarray}
 \frac{dt^2}{f} & = &  (1+\varepsilon) \frac{k\,l_s^2 dU^2}{U^2} ~. \label{dts}
\end{eqnarray}

Using \eqref{dts}, the induced metric on this timelike surface can be written as
\begin{eqnarray}
ds^2 =  \frac{1}{f^2}(-dt^2+dx^2) + \frac{k\,l_s^2}{U^2} dU^2 =   -\frac{\varepsilon\,k\,l_s^2}{U^2} dU^2 + \frac{1}{f^2} dx^2~.
\end{eqnarray}
The negative sign in the first term clearly indicates that this is a timelike surface. The unit outward normals to the surface \eqref{GHYbdry} are,
\begin{equation}
n^t = - \frac{t-T}{\sqrt{(1+\varepsilon)^2A^2(U)- \frac{(t-T)^2}{k\,l_s^2} }} \frac{\sqrt{f(U)}}{\sqrt{k}l_s},\ \ \ n^U =  - \frac{(1+\varepsilon) A(U) }{\sqrt{(1+\varepsilon)^2A^2(U)- \frac{(t-T)^2}{k\,l_s^2} }} \frac{U}{\sqrt{k}l_s}~, \ \ \  n^x=0~.
\end{equation}
The trace of the extrinsic curvature 
\begin{eqnarray}
\begin{split}
K \equiv \nabla_L\,  n^L &= \partial_L \, n^L + \Gamma^L_{LM}\,  n^M  
=  \partial_t n^t + \partial_U n^U + \Gamma^L_{L U} n^U~. \label{K}
\end{split}
\end{eqnarray}
takes the form
\begin{eqnarray}
K & = & \frac{2}{\sqrt{\varepsilon} \sqrt{k} l_s (1+\lambda k U^2) }~.
\end{eqnarray}

Thus the GHY term for this surface in the null limit ($\varepsilon\rightarrow 0$) is
\begin{eqnarray}
\begin{split}
S^{\partial WdW}_{GHY} &= \lim_{\varepsilon\rightarrow0} \,\,\frac{
1}{8\pi G_N}\int d^2X  e^{-2(\Phi-\Phi_0)} \sqrt{-\gamma} \,\,\left[K - 4 n^M \partial_M \Phi\right]  \\
& =  \frac{L_x k}{4 \pi G_N} \int_0^{l_s/\epsilon} dU\,U \,\frac{2 \lambda +\frac{1}{k U^2}}{\sqrt{f}}  
=\frac{cL_x}{12\pi^2 \beta_h}\frac{\beta_H^2}{\epsilon^2}\sqrt{1+4\pi^2\frac{\epsilon^2}{\beta_H^2}}~.
\end{split}
\label{GHYnull0}
\end{eqnarray}
In the UV, $S^{\partial WdW}_{GHY} $ diverges as
\begin{equation}
\lim_{\epsilon/\beta_H\ll 1}S^{\partial WdW}_{GHY} = \frac{cL_x}{12\pi^2 \beta_h}\frac{\beta_H^2}{\epsilon^2}+\frac{cL_x}{6\beta_H}+O\left(\epsilon^2/\beta_H^2\right)~ . \label{GHY2}
\end{equation}
 In the IR one can write
 \begin{equation}
\lim_{\epsilon/\beta_H\gg 1}S^{\partial WdW}_{GHY} = \frac{cL_x}{6\pi \beta_H}\frac{\beta_H}{\epsilon}+\frac{cL_x}{48\pi^3\beta_H}\left(\frac{\beta_H}{\epsilon}\right)^3+O\left(\beta_H^4/\epsilon^4\right) ~. \label{GHY2}
\end{equation}

\subsubsection{Full Action complexity at zero temperature} 

\begin{figure}[h]
    \centering
    \includegraphics[width=.5\textwidth]{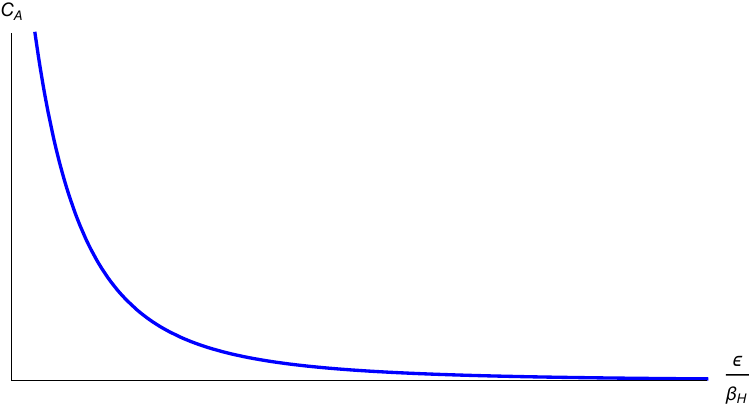}
    \caption{$C_\mathcal{A}(\epsilon/\beta_H)$ vs $\epsilon/\beta_H$ at $T=0$.}
    \label{img2}
\end{figure}

\begin{figure}[h]
    \centering
    \includegraphics[width=.5\textwidth]{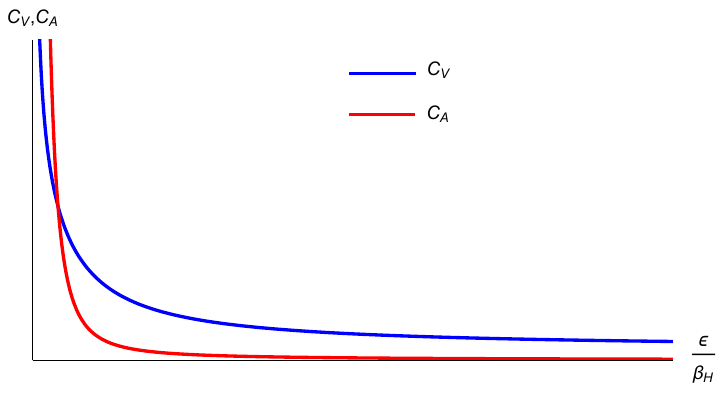}
    \caption{Comparison between $C_V$ and $C_\mathcal{A}$ at zero temperature. For large $\epsilon/\beta_H$, the action complexity decays much faster than volume complexity.}
    \label{img3}
\end{figure}

Putting together all the pieces, the full on-shell action over the WdW patch  is obtained by summing over the contributions \eqref{s1},\eqref{s2},\eqref{s3},\eqref{s4},\eqref{SP} and \eqref{GHYnull0}. The full action complexity \eqref{caa} thus obtained is presented in figure \ref{img2}.
In the UV linear dilaton regime (\ie\ when $\epsilon/\beta_H\ll1$), the action complexity (obtained by summing over the contributions \eqref{SR},\eqref{SPhi}, \eqref{Scc},\eqref{SP},\eqref{SH}, and \eqref{GHY2},) diverges as 
\begin{equation}\label{CA}
C_\mathcal{A} = \frac{L_x c}{ 3\pi^2 \beta_H} \left[\frac{\beta_H^2}{2\pi \epsilon^2}-2\pi \log\left(\frac{\beta_H}{\pi\epsilon}\right)+ \pi +O\left(\frac{\epsilon}{\beta_H}\right)\right]~.
\end{equation}
Comparison of \eqref{CA} with the volume complexity expression \eqref{CV@0T} reveals that the leading divergence structure (\ie\ the quadratic divergent term) and the constant term in both cases are identical. The subleading logarithmic divergences differ by a negative sign.
In the IR  (\ie\ when $\epsilon/\beta_H\gg1$) the action complexity takes the form
\begin{eqnarray}
\lim_{\epsilon/\beta_H\gg 1}C_\mathcal{A}=\frac{cL_x}{18\pi^3\beta_H}\left(\frac{\beta_H}{\epsilon}\right)^2+O\left(\beta_H^5/\epsilon^5\right)~. \label{CA_IR_0T}
\end{eqnarray}
Thus in pure $AdS_3$ the action complexity goes to zero. This is in precise agreement with the analysis performed in \cite{Reynolds:2016rvl}. Unlike the volume complexity, the action complexity in $\mathcal{M}_3$ decreases much faster. A comparison between volume complexity and action complexity in $\mathcal{M}_3$ is given in figure \ref{img3}.
Similar to the volume complexity, the action complexity diverges in the UV (\ie\ when $\epsilon/\beta_H\to 0$). Then as $\epsilon/\beta_H$ increases, the action complexity decreases (much faster than volume complexity) monotonically eventually going to 0 in the deep IR.

 \section{Holographic Complexity in $\mathcal{M}_3$ at finite temperature}  \label{finiteT}
\begin{figure}
\begin{center}
\begin{tikzpicture}[scale=1.3, transform shape]
\tikzset{middlearrow/.style={
        decoration={markings,
            mark= at position 0.5 with {\arrow{#1}} ,
        },
        postaction={decorate}
    }
}
 \draw [fill=pink!30]  (1.5,.5) -- (0,2)--(-1.5,.5)--(0,-1)--(1.5,.5); 
  \draw [black,-](2,0) -- (0,2)--(-2,0)--(0,-2)--(2,0);
  \draw [black,-](2,2) -- (0,4)--(-2,2)--(0,0)--(2,2);
   \draw [black,-](1,1) -- (-1,-1);
   \draw [black,-](-1,1) -- (1,-1);
   \draw [black,-](0,2) -- (1,3);
      \draw [black,-](0,2) -- (-1,3);
       \draw [red,decorate,decoration=zigzag](1,1) -- (1,3);
         \draw[red,decorate,decoration=zigzag](-1,1) -- (-1,3);
       \draw (1.2,2) node [rotate=90] {\tiny{singularity}};
              \draw (-1.2,2) node [rotate=90] {\tiny{singularity}};
              \draw (.55,.35) node [rotate=45] {\tiny{$U=U_T$}};
                \draw (-.55,.35) node [rotate=-45] {\tiny{$U=U_T$}};
                \draw (.52,1.3) node [rotate=-45] {\tiny{$U=0$}};
                 \draw (-.52,1.3) node [rotate=45] {\tiny{$U=0$}};
                 \draw (0,-.5) node {\scriptsize{$IV$}};
                 \draw (.8,.2) node {\scriptsize{$I$}};
                  \draw (-.8,.2) node {\scriptsize{$III$}};
                  \draw (0,1) node {\scriptsize{$II$}};
                    \draw (-1.9,.5) node {\tiny{$t=t_L$}};
                    \draw (1.9,.5) node {\tiny{$t=t_R$}};
 \end{tikzpicture}
\caption{Penrose diagram of the eternal $\mathcal{M}_3$ black hole geometry with the Wheeler-deWitt (WdW) patch shaded in pink for the boundary time $t_L$ and $t_R$.}
  \label{fig:finiteT M3 WdW}
 \end{center}
\end{figure}
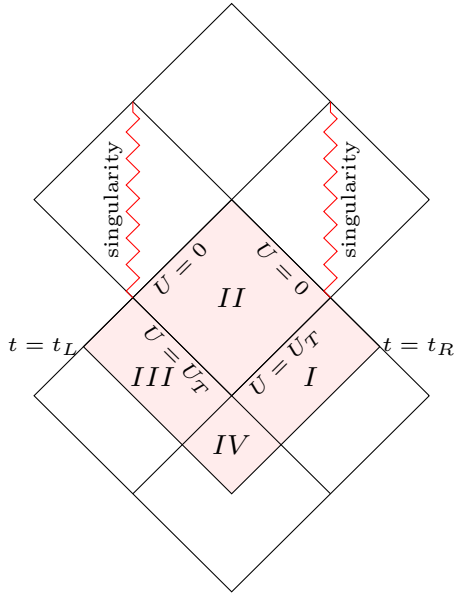

 In this section, we compute the holographic complexity for  LST at finite temperature. Our main aim is to look for new exotic divergence structures which do not arise in the zero temperature case and are endemic to finite temperatures exclusively. Although we have a good idea of what kind of finite temperature corrections one generates for complexity of \emph{local} quantum field theories and there we can rule out appearance of such exotic new divergences for finite temperatures, there is hardly such intuition for the case of nonlocal quantum field theories such as LST. In particular, we will be content by computing the action complexity as the integrals that can be performed numerically very easily without any approximations. Volume complexity on the other hand is a different story, the equations for the maximal volume slice are nonlinear and we could hope to solve (even numerically) perturbatively only in simple limits such as high temperatures or low temperatures. Instead of making such simplifying assumptions, we have decided to compute the action complexity exactly and evaluate the integrals numerically.  For this finite temperature case one has to use the finite temperature $\mathcal{M}_3$ background  \eqref{M3T}.  An important thing to note is the geometry here is that of the two-sided eternal  hole with four quadrants - right ($I$), future ($II$), left ($III$) and past ($IV$) wedges. The future wedge ($II$) is the region between the inner and the outer horizons. The Penrose diagram of the finite temperature $\mathcal{M}_3$ spacetime with the WdW patch is displayed in figure \ref{fig:finiteT M3 WdW}. Although we denote the four wedges of the eternal black hole by $I,II,III,IV$ precisely in the sense discussed above, in the discussion that follows, we will refer to the four section obtained by taking an intersection of the WdW patch with the full two-sided eternal hole  as regions $I,II,III,IV$.

So in the zero temperature limit of this two-sided geometry, one will get $twice$ the action complexity value for that of the single sided zero temperature geometry. The Wheeler-deWitt patch for the eternal geometry is anchored at the Schwarzschild times, $t_R$ in the right quadrant and $t_L$ in the left quadrant. Of course in terms of boundary time coordinate, the left quadrant time is then, $-t_L$. We also consider the case when $t_L=-t_R =t $ is very large, since in this case the past wedge, quadrant $IV$ pinches off and its contribution to the complexity the vanishes. Also it is worth mentioning that since the metric in left and right wedges $I$ and $III$ are time-independent and  has reflection symmetry around $t=0$, the complexity contributions from the left and right are identical and independent of  $t_R$ or $t_L$. The action complexity contributions at finite temperature (two sided $\mathcal{M}_3$ black hole) are worked out in the following subsections, first the contributions from the bulk (volume) of the WdW patch followed by contributions from the surface/edges of the WdW patch. The results are then plotted in the figure \ref{img4}. We find the finite temperature complexity qualitatively displays similar monotonic behavior as a function of  $\epsilon/ \beta_H$ and that there are no new exotic divergence structures appearing up to second order in finite temperature corrections (\ie\ $O\left(U_T^2\right)$) in action complexity (see Appendix \ref{appA}).

\subsection{Action complexity at finite temperature}

\subsubsection{Bulk terms for finite temperature action complexity}

The bulk action \eqref{bulk action} consists of four types of contributions, namely from the Ricci scalar term, from the cosmological constant term, the dilaton kinetic term and the NS-NS H-field strength term. We will write the metric in infalling null coordinate $v$ and radial coordinate $r$ which are well defined in the quadrants $I$ and $II$ (see figure \ref{fig:finiteT M3 WdW}). In terms of these the (string frame) metric in quadrants $I$ and $II$ looks like
\begin{equation}
ds^2 = -\frac{f_1}{f} dv^2 + \frac{2\sqrt{k}\,l_s}{\sqrt{f}\,U} dv\,dU +\frac{dx^2}{f}~.
\end{equation}
The $v$ coordinate is related to the Schwarzschild coordinates, $t,U$ by the relation:
\begin{equation}
v= t + U_{\ast}~,
\end{equation}
where the (UV regularized) tortoise coordinate, $U_*(U)$ is defined by 
\begin{equation}
U_{\ast}=\left\{ \begin{array}{c}
\int_{\frac{l_{s}}{\epsilon}}^{U}dU'\frac{\sqrt{k}l_{s}}{f_{1}(U')}\frac{\sqrt{f(U')}}{U'}, \ \ \text{ region }I~,\\
\int_{0}^{U}dU'\frac{\sqrt{k}l_{s}}{-f_{1}(U')}\frac{\sqrt{f(U')}}{U'}, \ \text{ region }II.
\end{array}\right. \label{eq: tortoise}
\end{equation}

Of course these coordinates do not cover the left wedge $III$ or the past region $IV$. However, since the metric in region $III$ is time-independent (and time-reflection symmetric), it turns out that the contribution from $III$ is exactly equal to that of region $I$. As mentioned before we are looking at large / late times, \ie\ $t_R=-t_L = t \rightarrow \infty$, in this limit the wedge $IV$ pinches off and there is no contribution from it.  Here are the list of the bulk term contributions to the action complexity.\\

 \noindent  \textbf{The Ricci scalar term:}
 
 The Ricci scalar term in the supergravity action in region $I$ contributes
 \begin{eqnarray}
 	\begin{split}
 		S^I_R&=\frac{L_x}{16\pi\,G_N} \int^{\frac{l_s}{\epsilon}}_{U_T} dU\int_{t_R+2U_\ast(U)}^{t_R}dv \sqrt{-g}e^{-2(\Phi-\Phi_0)} R\\
 		&=\frac{L_xk^{\frac{1	}{2}}}{8\pi\,G_N} \int^{\frac{l_s}{\epsilon}}_{U_T} dU U \left(\frac{2 U_T^2 k \lambda  \left(2 k \lambda  U^2-5\right)+8 k \lambda  U^2-6}{ \left(k \lambda  U^2+1\right)^2}\right)\left(\int^{\frac{l_s}{\epsilon}}_{U}dU'\frac{\sqrt{\lambda k U'^2+1}}{(U'^2-U_T^2)}\right)~. 
 \end{split}
  \end{eqnarray}
 Owing to the symmetry between region $I$ and $III$ in the Penrose diagram, the region $III$ integrals give same contribution as region $I$ with just the change $t_R\Longleftrightarrow t_L$ interchange. Since, anchorage time does not feature in the integrals involving regions outside the outer horizon, we simply the exact same contribution from the region $III$. Therefore, 
 \begin{eqnarray}
 	S_R^{III}=S_R^{I}~.
 \end{eqnarray}

Next, the contribution coming from region II is given by 
\begin{eqnarray}
 	\begin{split}
 	S^{II}_R&=\frac{L_x}{16\pi\,G_N} \int_{0}^{U_T} dU\int_{t_L+2U_\ast(U)}^{t_R}dv \sqrt{-g}e^{-2(\Phi-\Phi_0)} R \\
 	&	 =\frac{L_xk^{\frac{3}{2}}l_s}{16\pi\,G_N} \int_{0}^{U_T} dU U \left(\frac{2 U_T^2 k \lambda  \left(2 k \lambda  U^2-5\right)+8 k \lambda  U^2-6}{k l_s^2 \left(k \lambda   U^2+1\right)^2}\right)\left(t_R-t_L-2l_s\int_{0}^{U}dU'\frac{\sqrt{\lambda k U'^2+1}}{(U'^2-U_T^2)}\right) ~.
  \end{split}
  \end{eqnarray}
The contribution from $S^{II}_R$ trivially goes to zero in the limit $U_T\to 0$. In the late times limit, there is no contribution from region $IV$ since it gets pinched off.\\

 \noindent  \textbf{The cosmological constant term:} 
 
This term is particularly simple since it is proportional to the volume of the WdW patch. The contribution to the onshell action  from regions outside the horizons (\ie\ region $I, III$) is 
\begin{eqnarray}
\begin{split}
 	S^I_{\Lambda}&=\frac{4 L_x}{16\pi kl_s^2\,G_N} \int^{\frac{l_s}{\epsilon}}_{U_T} dU\int_{t_R+2U_\ast(U)}^{t_R}dv  \sqrt{-g}e^{-2(\Phi-\Phi_0)}\\
 	&=\frac{4 L_x k^{\frac{1}{2}} }{16\pi l_s\,G_N} \int^{\frac{l_s}{\epsilon}}_{U_T} dU U \left(-2l_s\int_{\frac{l_s}{\epsilon}}^{U}dU'\frac{\sqrt{\lambda k U'^2+1}}{(U'^2-U_T^2)}\right)~.
	\end{split}
\end{eqnarray}
As argued before $S^{III}_\Lambda=S^{I}_\Lambda$. 
 The contribution to the volume of the WdW patch from inside the horizon region namely region $II$ is given by
 \begin{eqnarray}
 S^{II}_{\Lambda} &=& \frac{4 L_x k^{\frac{1}{2}} }{16\pi l_s G_N}\int^{U_T}_{0} dU U \left(t_R-t_L-2l_s\int_{0}^{U}dU'\frac{\sqrt{\lambda k U'^2+1}}{(U'^2-U_T^2)}\right)~.
 \end{eqnarray}
 As expected the $S^{II}_{\Lambda} $ vanishes in the limit $U_T \to 0$ and for large $t_R$, $S^{IV}_\Lambda\to 0$.\\

 \noindent  \textbf{The Dilaton kinetic term:}
 
 The dilaton kinetic term in the supergravity action coming from region $I$ is given by:
 \begin{eqnarray}
\begin{split}
 	S^I_{\Phi}&=\frac{4L_x}{16\pi\,G_N} \int^{\frac{l_s}{\epsilon}}_{U_T} dU \int_{t_R+2U_\ast(U)}^{t_R}dv\sqrt{-g}e^{-2(\Phi-\Phi_0)}g^{\mu\nu}\partial_\mu\Phi\partial_\nu\Phi \\
 	&=\frac{L_xk^{\frac{5}{2}}\lambda^2}{2\pi\,G_N} \int^{\frac{l_s}{\epsilon}}_{U_T} dU 	\left(\frac{ U^5}{\left(k \lambda   U^2+1\right)^2}-\frac{ U^3U_T^2}{\left(k \lambda   U^2+1\right)^2}\right) \left(\int^{\frac{l_s}{\epsilon}}_{U}dU'\frac{\sqrt{\lambda k U'^2+1}}{(U'^2-U_T^2)}\right)~.
	\end{split}
 \end{eqnarray}
 The contribution from the region $III$ is same is that of region $I$ namely $S^{III}_{\Phi}=S^I_{\Phi}$.
 
  The contributions from the region $II$ \ie\ inside the horizon in this case is
\begin{eqnarray}
 		S^{II}_{\Phi}=\frac{4L_xk^{\frac{3}{2}}l_s}{16\pi\,G_N} \int_{0}^{U_T} dU U	\left(\frac{k^2 \lambda ^2 U^2 \left(U^2-U_T^2\right)}{l_s^2\left(k \lambda   U^2+1\right)^2}\right) \left(t_R-t_L-2l_s\int_{0}^{U}dU'\frac{\sqrt{\lambda k U'^2+1}}{(U'^2-U_T^2)}\right)~.\nonumber \\
\end{eqnarray}
 As a check one can see that $S^{II}_{\Phi}\to 0$ as $U_T\to 0$ and for late time $t_R$, $S^{IV}_\Phi\to 0$. \\

 \noindent  \textbf{The Kalb-Ramond term:}\\
The contribution to action complexity from the Kalb-Ramond term in region $I$ is given by: 
\begin{eqnarray}
\begin{split}
 	S^I_{H}&=-\frac{L_x}{12\times 16\pi\,G_N} \int^{\frac{l_s}{\epsilon}}_{U_T} dU \int_{t_R+2U_\ast(U)}^{t_R}dv \sqrt{-g}e^{-2(\Phi-\Phi_0)}H^2  \\
 	&= -\frac{L_x}{4\pi G_N\,k^{\frac{3}{2}}}  \int_{U_T}^{l_s/\epsilon} \frac{dU}{U^3\,f^2}\left(\int_{\frac{l_s}{\epsilon}}^{U}dU'\frac{\sqrt{\lambda k U'^2+1}}{(U'^2-U_T^2)}\right)~.
	\end{split}
\end{eqnarray}

The contribution from the region $II$ \ie\ interior to the future horizon is,
 \begin{eqnarray}
 	S_H^{II}		= \frac{L_x}{8\pi G_N\,k^{\frac{3}{2}}l_s}  \int_{0}^{U_T} \frac{dU}{U^3\,f^2}\left(t_R-t_L-2l_s\int_{0}^{U}dU'\frac{\sqrt{\lambda k U'^2+1}}{(U'^2-U_T^2)}\right)~.
  \end{eqnarray}
Again, predictably this inside horizon contribution vanishes in the zero temperature limit.  Finally, the contribution from region $III$ is identical to that of region $I$ \ie\ $S^{III}_H=S^{I}_{H}$. For large $t_R$, $S^{IV}_H\to 0$.

\subsubsection{GHY term for the null boundaries of the WdW patch}

Let's first consider the right boundaries of the null $WdW$ patch
defined by the equations 
\begin{eqnarray}
v=t_{R}\qquad(\text{future}) \ \ \ \& \  \ \  v-2U_{\ast}=t_{R}\qquad(\text{past})~,
\end{eqnarray}
where $U_{\ast}$ is the tortoise coordinate for the outside horizon
region (region $I$) \eqref{eq: tortoise}. In region $I$, these two null boundaries can
be combined and deformed into a continuous timelike surface defined
by equation
\begin{equation}
\frac{\left(t-t_{R}\right)^{2}}{kl_{s}^{2}}-\left(1+\varepsilon\right)A^{2}(U)=0~,\quad \text{where }\ \ A(U)=\int_{l_{s}/\epsilon}^{U}dU'\frac{\sqrt{f(U')}}{U'f_{1}(U')}~.\label{eq: Deformation surface for right null boundary of WdW patch}
\end{equation}
where $\varepsilon$ is the deformation parameter which when sent to zero, takes the above timelike surface into a pair of null surfaces.\footnote{This ``null-to-timelike" deformation parameter $\varepsilon$ is in principle independent of the UV regulator $\epsilon$, but can be chosen, without inconsistency to be equal to $\epsilon$ (see \eg\ \cite{Bolognesi:2018ion}).} Note that by definition, $A(U)<0$. Let us denote this surface by $\Gamma$.
The induced metric on the deformation surface $\Gamma$ is given by 
\begin{eqnarray}
ds^{2}=-\varepsilon\frac{kl_{s}^{2}}{f_{1}U^{2}}dU^{2}+\frac{dx^{2}}{f}~.
\end{eqnarray}
Hence,
\begin{equation}
dx\;dU\;\sqrt{-\gamma}\;e^{-2\left(\Phi-\Phi_{0}\right)}=\sqrt{\varepsilon}k^{3/2}l_{s}dx\;dU\;U\sqrt{\frac{f}{f_{1}}}\label{eq: measure part for GHY outside horizon}
\end{equation}
where $\gamma$ denotes the determinant of the induced metric on the surface $\Gamma$.
Next, we compute the trace of the extrinsic curvature of the surface
\eqref{eq: Deformation surface for right null boundary of WdW patch}.
The components of the unit outward normal are
\begin{eqnarray}
\begin{split}\label{norvect}
&n^{t}=-\frac{t-t_{R}}{\sqrt{k}\;l_{s}\sqrt{\left(1+\varepsilon\right)^{2}A^{2}-\frac{\left(t-t_{R}\right)^{2}}{kl_{s}^{2}}}}\sqrt{\frac{f}{f_{1}}}~,\\
& n^{U}=-\frac{\left(1+\varepsilon\right)U\:A}{\sqrt{k}\;l_{s}\sqrt{\left(1+\varepsilon\right)^{2}A^{2}-\frac{\left(t-t_{R}\right)^{2}}{kl_{s}^{2}}}}\sqrt{f_{1}}~,\\
&n^{x}=0~.
\end{split}
\end{eqnarray}
Using the above information one can write
\begin{eqnarray}
\left.\left(K-4n^{U}\partial_{U}\Phi\right)\right|_{\Gamma}=2\frac{1+2\lambda kU^{2}}{1+\lambda kU^{2}}\frac{\sqrt{f_{1}}}{\sqrt{k}\;l_{s}\sqrt{\varepsilon}}+\frac{1}{\sqrt{k}\;l_{s}\sqrt{\varepsilon}}\frac{1}{\sqrt{f_{1}}}\frac{U_{T}^{2}}{U^{2}}~.
\end{eqnarray}
Thus the GHY term contribution from the right null boundary in region $I$ is given by
\begin{eqnarray}\label{ghyone}
\begin{split}
S_{GHY}^{\partial WdWI} & =\frac{1}{8\pi G_{N}}\int d^{2}X\;\sqrt{-\gamma}\;e^{-2\left(\Phi-\Phi_{0}\right)}\left.\left(K-4n^{U}\partial_{U}\Phi\right)\right|_{\Gamma}\\
 & =\frac{L_{x}\:k}{4\pi G_{N}}\;\int_{U_{T}}^{ls/\epsilon}dU\;U\frac{2\lambda+\frac{1}{kU^{2}}}{\sqrt{f}}+\frac{L_{x}\:k}{8\pi G_{N}}\;U_{T}^{2}\int_{U_{T}}^{ls/\epsilon}dU\frac{\sqrt{f}}{Uf_{1}}~.
 \end{split}
\end{eqnarray}
Evidently, when one sets $U_{T}=0$, this reduces precisely to the
GHY contribution for the zero temperature case for the right null
boundary of $WdW$ patch \eqref{GHYnull0},
One can exactly evaluate the integral \eqref{ghyone} to obtain
\begin{eqnarray}
\begin{split}
S_{GHY}^{\partial WdWI} & =\frac{L_{x}\:k}{4\pi G_{N}}\left(\sqrt{\lambda}\frac{l_{s}^{2}}{\epsilon^{2}}+\frac{1}{2k\sqrt{\lambda}}-U_{T}^{2}\sqrt{f\left(U_{T}\right)}\right) \\
 &+ \frac{L_{x}\:k}{8\pi G_{N}}\;U_{T}^{2}\left(\sqrt{\lambda}\ln\left(\frac{2\sqrt{\lambda k}l_{s}}{\epsilon}\right)+\sqrt{f\left(U_{T}\right)}\ln\left(\sqrt{1+\lambda kU_{T}^{2}}-\sqrt{\lambda k}U_{T}\right)\right) \\
 &- \frac{L_{x}\:k}{8\pi G_{N}}\;U_{T}^{2}\left(\sqrt{\lambda}\sinh^{-1}\left(\sqrt{\lambda k}U_{T}\right)-\frac{\sqrt{f\left(U_{T}\right)}}{2}\lim_{U\rightarrow U_{T}^{+}}\ln\left(\frac{2U_{T}\left(1+\lambda kU_{T}^{2}\right)}{U-U_{T}}\right)\right)~.\label{eq: GHY term from the right null boundary of WdW patch in region I}
 \end{split}
\end{eqnarray}

Next, we evaluate the GHY contribution from the part of the right null
boundary of the $WdW$ patch from within the horizon \ie\ in region
$II$. In this case it is simpler to work with the deformed timelike
surface
\begin{eqnarray}
t_{R}-t=\left(1-\delta\right)\sqrt{k}\;l_{s}\;B(u)~,\qquad \text{where } \ \ \ \ \  B(u)\equiv\int_{0}^{U}\frac{dU'\sqrt{f(U')}}{U'\:\left(-f_{1}(U')\right)}~.
\end{eqnarray}
The induced metric on the right null boundary of the $WdW$ patch is 
\begin{eqnarray}
ds^2=-2\delta\:\frac{kl_{s}^{2}}{U^{2}\left(-f_{1}\right)}dU^{2}+\frac{dx^{2}}{f}~.
\end{eqnarray}
The unit outward normal is given by
\begin{eqnarray}
n^{t}=\frac{1}{\sqrt{2\delta}}\sqrt{\frac{f}{-f_{1}}},\qquad n^{U}=\frac{\left(1-\delta\right)}{\sqrt{2\delta}}\frac{U\sqrt{-f_{1}}}{\sqrt{k}\;l_{s}},\qquad n^{x}=0~.\label{unit normal right}
\end{eqnarray}
The full integrand of the GHY term is
\begin{eqnarray}
K-4n^{L}\partial_{L}\Phi=\frac{1}{\sqrt{2\delta}\sqrt{k}\;l_{s}}\left(-\frac{1}{\sqrt{-f_{1}}}\frac{U_{T}^{2}}{U^{2}}+2\left(\frac{1+2\lambda kU^{2}}{1+\lambda kU^{2}}\right)\sqrt{-f_{1}}\right). \label{t-def right}
\end{eqnarray}
Thus, the GHY term contribution from the right null boundary of the
WdW patch in region $II$ is
\begin{eqnarray}
S_{GHY}^{\partial WdW_{R}II} & =&\frac{1}{8\pi G_{N}}\int d^{2}X\;\sqrt{-\gamma}\left(K-4n^{L}\partial_{L}\Phi\right)\nonumber  \\
 & =&-\frac{L_{x}k}{8\pi G_{N}}U_{T}^{2}\left(\frac{\sqrt{f(U_{T})}}{2}\lim_{U\rightarrow U_{T}^{-}}\ln\left(\frac{U_{T}\left(1+\lambda kU_{T}^{2}\right)}{U_{T}-U}\right)-\sqrt{\lambda}\sinh^{-1}\left(\sqrt{\lambda k}U_{T}\right)\right)\nonumber \\
 &&+\frac{L_{x}k}{4\pi G_{N}}U_{T}^{2}\sqrt{f(U_{T})}~. \label{eq: GHY term contribution from the right null boundary of WdW patch in region II}
\end{eqnarray}

Thus summing the contributions from both outside and inside the horizon,
\ie\ \eqref{eq: GHY term from the right null boundary of WdW patch in region I}
and \eqref{eq: GHY term contribution from the right null boundary of WdW patch in region II}
we obtain the GHY type contributions to action from the right null
boundary of the $WdW$ patch as
\begin{eqnarray}
\begin{split}
S_{GHY}^{\partial WdW_{R}}&=\frac{L_{x}\:k}{8\pi G_{N}}\;U_{T}^{2}\left(\sqrt{\lambda}\ln\left(\frac{2\sqrt{\lambda k}l_{s}}{\epsilon}\right)+\sqrt{f\left(U_{T}\right)}\ln\left(\sqrt{1+\lambda kU_{T}^{2}}-\sqrt{\lambda k}U_{T}\right)\right)\\
&\qquad\qquad\qquad\qquad\qquad\qquad\qquad\qquad\qquad+\frac{L_{x}\:k}{4\pi G_{N}}\left(\sqrt{\lambda}\frac{l_{s}^{2}}{\epsilon^{2}}+\frac{1}{2k\sqrt{\lambda}}\right)~.\label{eq: GHY term contribution from the right null boundary of the WdW patch}
\end{split}
\end{eqnarray}

\subsubsection{Joint contributions for the intersection of null boundaries of WdW patch}

Here we compute the contribution to the action (complexity) supported on the joint or edge along which the null boundaries of the WdW patch intersects. The future boundaries of the WdW patch are along the inner horizon, $U=0$ (refer to figure \ref{fig:finiteT M3 WdW}). Since we have deformed the null boundaries of the WdW to timelike and we take the null limit only at the very end, we are considering a joint of two timelike surfaces along $U=0$. The right future null boundary has been deformed to a timelike surface \eqref{t-def right} with the unit outward normal given in \eqref{unit normal right}. Analogously the left future null boundary of the WdW patch, namely, $t-t_L=U_*$ can be deformed to timelike,
\begin{equation}
t-t_L = (1-\delta) \sqrt{k} l_s B(U)
\end{equation}
where  $B(U)$ has already been defined in eq. (\ref{t-def right}). The unit outward normal for this timelike deformed boundary is,
\begin{equation}
\bar{n}_t = \frac{1}{\sqrt{2\delta}} \sqrt{\frac{-f_1}{f}},\,\,\,\,\bar{n}_U = -\frac{1-\delta}{\sqrt{2\delta}} \frac{\sqrt{k}l_s}{U\sqrt{-f_1}}, \,\,\,\,\bar{n}_x=0. \label{unit normal left}
\end{equation}
From the expression of the unit outward normals \eqref{unit normal right} and \eqref{unit normal left}, it is evident that,  $n.\bar{n}=1$ and hence
\begin{equation}
\ln\left| n.\bar{n}\right| = 0.
\end{equation}
Thus the joint contribution (evaluated in the Einstein frame) vanishes,
\begin{eqnarray}
S^{U=0}_{\cap\partial WdW} & = & \frac{1}{8\pi G_N} \int dx \lim_{U\rightarrow0} \left(\sqrt{\tilde{g}_{xx}} \ln\left| n.\bar{n}\right|\right) =0.
\end{eqnarray}
because $\lim_{U\rightarrow0} \sqrt{\tilde{g}_{xx}} \rightarrow\frac{\sqrt{k}U}{g_s^2}$. Here $\tilde{g}$ denotes the Einstein frame metric, $\tilde{g}=e^{-4(\Phi-\Phi_0)} g$.

\subsubsection{Full Action complexity at finite temperature}

Thus the full action complexity for the finite temperature case in the late time limit, is given by gathering together contributions from regions $I,II, \& III$  (with the contributions from region $III$ being identical to those from region $I$),
\begin{equation}
 C_{\mathcal{A}} = \frac{1}{\pi \hbar} \left(2S_R^{I}+S_R^{II}+2 S^I_{\Lambda} + S^{II}_{\Lambda}+ 2S^I_{\Phi}+S^{II}_{\Phi}+2S^{I}_H+S^{II}_{H}+S_{GHY}^{\partial WdW_{R}} + S_{GHY}^{\partial WdW_{L}} + S^{U=0}_{\cap\partial WdW} \right)~.
\end{equation}

\begin{figure}[h]
    \centering
    \includegraphics[width=.5\textwidth]{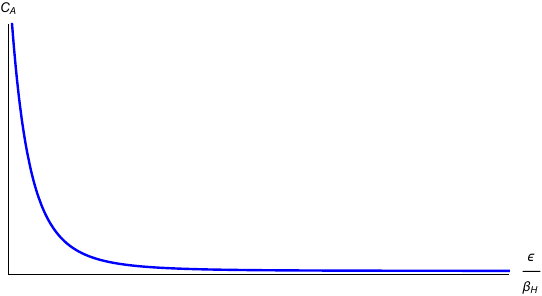}
    \caption{$C_\mathcal{A}(\epsilon/\beta_H)$ vs $\epsilon/\beta_H$ at finite temperature $(T_{bh}/T_H=0.1)$.}
    \label{img4}
\end{figure}
  Figure \ref{img4} shows the plot of action complexity at finite temperature as a function of $\epsilon/\beta_H$. As in the case of zero temperature, the action complexity monotonically decreases from the UV to the IR. In Appendix \ref{appA} we have performed the asymptotic analysis of the action complexity term by term perturbatively in finite temperature up to second order \ie\ $O\left(U_T^2\right)$ to extract the UV divergence structure. Turning on a finite temperature doesn't give rise to new exotic temperature dependent divergences (at least up to second order in $U_T^2$) that go away in the zero temperature limit at least up to second order in $U_T$. This is perhaps expected from the physical insight that the finite temperature introduces a horizon deep inside but does not change the asymptotic structure of the geometry and hence no new UV divergences are not expected to appear at finite temperature.
 
\section{Discussion \& Outlook} \label{DiscOut}

In this paper, we studied string theory in the background that interpolates between $AdS_3$ in the IR to flat spacetime with a linear dilaton in the UV both at zero \eqref{M3} and at finite temperature \eqref{M3T}. We studied holographic complexity using the $CV$ and $CA$ conjecture in this background and investigated the effects of non-locality of LST through the lenses of holographic complexity. Here is a summary of our findings:

\begin{itemize}

\item At zero temperature, both the volume and action complexities are UV divergent and hence manifestly regulator dependent. In the regime where the UV cutoff (lattice spacing) is shorter than the Hagedorn scale of the LST, the leading piece diverges \emph{quadratically} with the UV cutoff \eqref{CV@0T}. We identify this quadratic divergence as the characteristic signature of nonlocal nature of the LST. Modulo an overall factor ambiguity (which is well known in the literature) the leading divergences for both complexities (volume and action) agree and have the same sign. 

\item There are subleading logarithmic divergences in both volume complexity \eqref{CV@0T} and action complexity expressions \eqref{CA} which have the same magnitude but differ in sign. The universal coefficient \eqref{total dof} of this log divergent term can be interpreted as the total number of degrees of freedom in the LST  with the Hagedorn scale, $\beta_H$ treated as the lattice spacing.

\item In the opposite regime, \ie\ when the UV cutoff is much larger than the Hagedorn scale, the volume complexity expression expectedly reduces to that of a local field theory \ie\ having linear divergence (corresponding to a single spatial dimension) \eqref{CVir}. In fact this expression matches that of a CFT with the central charge equal to the Brown-Henneaux expression derived from a pure $AdS_3$ calculation. Similarly, in this limit the action complexity too reproduces the expected pure $AdS_3$ answer \eqref{CA_IR_0T} \cite{Reynolds:2016rvl, Carmi:2016wjl}.

\item At finite temperature we computed the action complexity since it can be computed exactly (numerically) without any approximations. The finite temperature complexity displays the same qualitative features as that of the zero temperature case, in particular it monotonically decrease with $\epsilon/\beta_H$. We do not find any new exotic divergences at finite temperature compared to the zero temperature case, at least perturbatively up to second order in finite temperature corrections.
\end{itemize}

The leading divergence of both the volume and the action complexity, at short distances (\ie\ $\epsilon/\beta_H\ll 1$), goes as inverse of the square of the short distance cutoff scale (\ie\ $C_{V,\mathcal{A}}\sim 1/\epsilon^{2}$).  A striking feature of the above fact is that LST visualized as a six-dimensional theory on NS5/M5 in \eg\ type IIA/M-theory, will also exhibit the same quadratic (leading) and logarithmic (subleading) divergences. This is due to the fact that the two-dimensional LST we are interested in can be thought of as a $T^4$ compactification of the six-dimensional LST. Such a six dimensional LST (in type IIA theory) flows to a fixed point in the IR, the so called six-dimensional $(2,0)$ SCFT. The complexity of this SCFT$_6$ (unlike CFT$_2$/SCFT$_2$) has a leading divergence that goes like $V_5/\epsilon^5$ \cite{ Carmi:2016wjl, Reynolds:2016rvl} (as opposed to $1/\epsilon$ in the case of CFT$_2$), where $V_5$ is the five dimensional spacial volume of the manifold on which the CFT$_6$ lives . It would be nice to have a more intuitive understanding of the universality (independence of dimension) of the divergence structure of complexity of LST in various dimensions.

For $\epsilon/\beta_H\ll 1$, the sub-leading divergence in both complexities (volume and action) turn out to be a log term. Once again, the presence of the log term is another signature of non-locality.  In fact the absolute value of the log term can be considered as an effective number of degrees of freedom, $cL_x/\beta_H$ of the system provided we treat $\beta_H$ as the lattice spacing of the theory. It would be interesting to understand a precise relationship between the coefficient of the log term and the regularized\footnote{We call it ``regularized" degrees of freedom because the actual number of degrees of freedom of LST is infinite \cite{Barbon:2008ut,Chakraborty:2018kpr}.} degrees of freedom of LST.  The coefficient of the log term comes with opposite signs in the volume and action complexity. It would also be nice to have a more physical understanding of this discrepancy.  

The analysis of holographic Wilson loop \cite{Chakraborty:2018aji}, holographic entanglement entropy \cite{Chakraborty:2018kpr,Asrat:2019end,Asrat:2020uib,Chakraborty:2020udr} and thermodynamics \cite{Chakraborty:2020xyz,Chakraborty:2020swe} in $\mathcal{M}_3$ naturally reveals the non-locality scale through some pathologies in the physical observables. For example, the free energy and the entropic c-function diverges as the RG scale approaches the non-locality scale of LST. The partition function in the thermodynamic limit develops a branch cut singularity as the temperature approaches the Hagedorn temperature of LST. In our analysis of holographic complexity, we didn't come across such pathologies. We believe that such a pathology will be encountered in the analysis of subregion complexity. It would be interesting to do the exercise of subregion complexity to verify this fact.

Since the $\mathcal{M}_3$-LST correspondence is novel, non-AdS/non-CFT case of holography, perhaps a more pressing exercise is to work out the bulk-boundary holographic dictionary ala GKPW as well as HKLL \cite{Hamilton:2005ju, Hamilton:2006az, Hamilton:2006fh}. We can expect some very interesting twists in the bulk-boundary maps in this case because such maps will reconstruct \emph{local} supergravity excitations in the bulk, from \emph{nonlocal} excitations of the LST in the boundary. In the usual AdS/CFT setting such local bulk reconstruction maps depend on the symmetry structure as well as key locality/microcausality properties of the boundary CFT correlators, \eg\ in the HKLL setup bulk locality in the longitudinal directions is a direct consequence of CFT locality in the longitudinal directions, and the nontrivial challenge there is to understand bulk locality in the emergent radial (holographic) direction. However in the case of LST, the field theory is \emph{nonlocal even in the longitudinal directions}. It will be interesting to identify which alternative properties of a nonlocal theory such as the LST plays that key role in emergence of the quasilocal bulk $\mathcal{M}_3$ in both radial as well longitudinal directions.

Recently, some progress has been made in understanding a solvable irrelevant deformation of a CFT$_2$ by a Lorentz symmetry breaking operator that goes in the name of $J\bar{T}$ deformation \cite{Guica:2017lia,Chakraborty:2018vja}. Single trace $J\bar {T}$ deformation has been studied in \cite{Chakraborty:2018vja,Apolo:2018qpq} with further generalizations studied in \cite{Chakraborty:2019mdf,Chakraborty:2020cgo}. It would be interesting to understand the holographic complexity in these more general setups. Since the presence of $J\bar{T}$ deformation breaks Lorentz invariance, it would be interesting to understand its effect on volume and action complexity.

\section*{Acknowledgements}

The authors would like to thank D. Kutasov for many valuable discussions and perceptive comments on the manuscript. The authors would also like to thank A. Giveon for pointing out a mistake in the previous version. The work of SC is supported by the Infosys Endowment for the study of the Quantum Structure of Spacetime. The work of SR is supported by the IIT Hyderabad
seed grant SG/IITH/F171/2016-17/SG-47. The work of GK is supported by a Senior Research Fellowship (SRF) from the Ministry of Education (MoE)\footnote{formerly the Ministry of Human Resource Development (MHRD)}, Govt. of India. SR thanks the Dept. of Theoretical Physics (DTP) of the Tata Institute of Fundamental Research (TIFR), Mumbai, India for their warm hospitality during which this work was conceived. SR also thanks Arpan Bhattacharyya for valuable discussions.

\appendix

\section{Perturbative analysis of divergences arising at finite temperature}\label{appA}

The general form of the total contribution coming from the integrals from the various regions of the WdW patch for the finite temperature case are typically of the following form
\begin{equation}
2\int_{U_T}^{\frac{l_s}{\epsilon}}dU\,f(U,U_T,\epsilon)+\int_{0	}^{U_T}dU g(U,U_T)
\end{equation}
Where, the first integral is contributed by both exterior regions and the last integral is for the Region $II$. We notice that from these integrals, only the first one contains the information about the asymptotic boundary region. We are hence forth interested in looking at only the first integral \ie\ from the exterior region.\\
Since, the zero temperature solution is already known to us, we will treat the finite temperature as the perturbation to the zero temperature. Therefore, we are only interested in the terms which comes from the corrections to the zero temperature. To do this, we do a Taylor series expansion of the term of the interest and arrive at:
\begin{multline}
	2\int_{U_T}^{\frac{l_s}{\epsilon}}dU\,f(U,U_T,\epsilon)\simeq 2\int_{0}^{\frac{l_s}{\epsilon}}dU\,f(U,0,\epsilon)+2U_T\Bigg(\frac{d}{dU_T}	\int_{U_T}^{\frac{l_s}{\epsilon}}dU\,f(U,U_T,\epsilon)\Bigg)_{U_T=0}\\+2U_T^2\left(\frac{d2}{d U_T^2}\int_{U_T}^{\frac{l_s}{\epsilon}}dU\,f(U,U_T,\epsilon)\right)_{U_T=0}
	\end{multline}
The zeroth order result in temperature is known to us, therefore the higher order corrections are:
	\begin{multline}
	-2U_T\,\left[\Bigg(f(U,U_T,\epsilon)\Bigg)_{U=U_T}\right]_{U_T=0}+2U_T\left[\Bigg(\int_{U_T}^{\frac{l_s}{\epsilon}}dU\frac{df(U,U_T,\epsilon)}{dU_T}\Bigg)_{U=U_T}\right]_{U\rightarrow0}-U_T^2\Bigg[\frac{d}{dU_T}f(U_T,U_T,\epsilon)\Bigg]_{U_T=0}\\-U_T^2\left[\Bigg(\frac{d}{dU_T}f(U,U_T,\epsilon)\Bigg)_{U=U_T}\right]_{U_T=0}+U_T^2\Bigg[\int^{l/\epsilon}_{U_T}dU\frac{d^2}{dU_T^2}f(U,U_T,\epsilon)\Bigg]_{U_T=0}.
\end{multline}
\\
 \noindent  \textbf{The Ricci scalar term \& the dilaton kinetic term:}
 
 The Ricci scalar term in the supergravity action in region $I$ contributes
 \begin{eqnarray}
 	\begin{split}
 		S^I_R&=\frac{L_x}{16\pi\,G_N} \int^{\frac{l_s}{\epsilon}}_{U_T} dU\int_{t_R+2U_\ast(U)}^{t_R}dv \sqrt{-g}e^{-2(\Phi-\Phi_0)} R\\
 		&=\frac{L_xk^{\frac{1	}{2}}}{8\pi\,G_N} \int^{\frac{l_s}{\epsilon}}_{U_T} dU U \left(\frac{2 U_T^2 k \lambda  \left(2 k \lambda  U^2-5\right)+8 k \lambda  U^2-6}{ \left(k \lambda  U^2+1\right)^2}\right)\left(\int^{\frac{l_s}{\epsilon}}_{U}dU'\frac{\sqrt{\lambda k U'^2+1}}{(U'^2-U_T^2)}\right)~. 
 \end{split}
  \end{eqnarray}
	\begin{equation}
		\begin{split}
	S^I_{\Phi}&=\frac{4L_x}{16\pi\,G_N} \int^{\frac{l_s}{\epsilon}}_{U_T} dU \int_{T_R+2U_*(U)}^{T_R}dv\sqrt{-g}e^{-2(\Phi-\Phi_0)}g^{\mu\nu}\partial_\mu\Phi\partial_\nu\Phi \\
	&=\frac{L_xk^{\frac{5}{2}}\lambda^2}{2\pi\,G_N} \int^{\frac{l_s}{\epsilon}}_{U_T} dU 	\left(\frac{ U^5}{\left(k \lambda   U^2+1\right)^2}-\frac{ U^3U_T^2}{\left(k \lambda   U^2+1\right)^2}\right) \left(\int^{\frac{l_s}{\epsilon}}_{U}dU'\frac{\sqrt{\lambda k U'^2+1}}{(U'^2-U_T^2)}\right)
\end{split}
\end{equation}
The finite temperature correction to the sum of Ricci scalar and the dilaton kinetic term is
\begin{equation}
\frac{3 k \sqrt{\lambda }\,U_T L_x }{2 \pi  G_N}\log \left(\frac{2 \sqrt{k \lambda } l_s}{\epsilon }\right)-\frac{ 3 k \sqrt{\lambda } L_x U_T^2}{2 \pi  G_N}\log \left(\frac{2 \sqrt{k \lambda } l_s}{\epsilon }\right)
\end{equation}
Where, the first order temperature correction comes from the Ricci term. At the level of second order correction, the Ricci and the dilaton kinetic contributions add up to give a logarithmic divergence.\\

 \noindent  \textbf{The cosmological constant term:} 
 
The contribution to the onshell action  from regions outside the horizons ( region $I, III$) is 
\begin{eqnarray}
\begin{split}
 	S^I_{\Lambda}&=\frac{4 L_x}{16\pi kl_s^2\,G_N} \int^{\frac{l_s}{\epsilon}}_{U_T} dU\int_{t_R+2U_\ast(U)}^{t_R}dv  \sqrt{-g}e^{-2(\Phi-\Phi_0)}\\
 	&=\frac{ L_x k^{\frac{1}{2}} }{2\pi \,G_N} \int^{\frac{l_s}{\epsilon}}_{U_T} dU U \left(\int^{\frac{l_s}{\epsilon}}_{U}dU'\frac{\sqrt{\lambda k U'^2+1}}{(U'^2-U_T^2)}\right)
	\end{split}
\end{eqnarray}
This term receives following asymptotic UV contribution from the finite temperature correction at $O(U_T^2)$
\begin{equation}
\simeq\frac{ -2k \sqrt{\lambda } L_x U_T^2}{\pi  G_N}\log \left(\frac{2 \sqrt{k \lambda } l_s}{\epsilon }\right)-\frac{2k^2 \lambda ^{3/2} l_s^2 L_x U_T^2}{3 \pi  \epsilon ^2 G_N}
\end{equation}


 \noindent  \textbf{The Kalb-Ramond term:}\\
The contribution to action complexity from the Kalb-Ramond term in region $I$ is given by: 
\begin{eqnarray}
\begin{split}
 	S^I_{H}&=-\frac{L_x}{12\times 16\pi\,G_N} \int^{\frac{l_s}{\epsilon}}_{U_T} dU \int_{t_R+2U_\ast(U)}^{t_R}dv \sqrt{-g}e^{-2(\Phi-\Phi_0)}H^2  \\
 	&= -\frac{L_x}{4\pi G_N\,k^{\frac{3}{2}}}  \int_{U_T}^{l_s/\epsilon} \frac{dU}{U^3\,f^2}\left(\int^{\frac{l_s}{\epsilon}}_{U}dU'\frac{\sqrt{\lambda k U'^2+1}}{(U'^2-U_T^2)}\right)~.
	\end{split}
\end{eqnarray}
The finite temperature correction
\begin{equation}
\simeq\frac{U_T^2 k^{5/2} \sqrt{\lambda } l_s L_x}{\pi  G_N}\log \left(\frac{2 \sqrt{k \lambda } l_s}{\epsilon }\right)
\end{equation}  
So up to second order in $U_T$ or finite temperature corrections there are no newer exotic type of divergences coming from the volume term of action complexity. 
The GHY term contribution \eqref{eq: GHY term contribution from the right null boundary of the WdW patch} is already exact to order $U_T^2$ and the correction is manifestly log-divergent. So no newer exotic divergences arise from the GHY term(s) either



\providecommand{\href}[2]{#2}\begingroup\raggedright\endgroup

\end{document}